\documentclass[12pt,preprint]{aastex631}

\usepackage{natbib}
\usepackage{graphicx}
\usepackage{epsfig} 
\usepackage{bm} 
\usepackage{amsmath}
\usepackage{amssymb} 
\usepackage{color} 
\usepackage{float}
\usepackage{hyperref}
\usepackage{wasysym}
\usepackage{wrapfig, xspace, lipsum, todonotes}

\def\flat{\textit{Fermi}-LAT\xspace}

\begin{document}

\title{Search for dark matter annihilation signals from unidentified \flat objects with H.E.S.S.}
\author{H.~Abdalla}\affiliation{University of Namibia, Department of Physics, Private Bag 13301, Windhoek 10005, Namibia}
\author{F.~Aharonian}\affiliation{Dublin Institute for Advanced Studies, 31 Fitzwilliam Place, Dublin 2, Ireland} \affiliation{Max-Planck-Institut f\"ur Kernphysik, P.O. Box 103980, D 69029 Heidelberg, Germany} \affiliation{High Energy Astrophysics Laboratory, RAU,  123 Hovsep Emin St  Yerevan 0051, Armenia}
\author{F.~Ait~Benkhali}\affiliation{Max-Planck-Institut f\"ur Kernphysik, P.O. Box 103980, D 69029 Heidelberg, Germany}
\author{E.O.~Ang\"uner}\affiliation{Aix Marseille Universit\'e, CNRS/IN2P3, CPPM, Marseille, France}
\author{C.~Arcaro}\affiliation{Centre for Space Research, North-West University, Potchefstroom 2520, South Africa}
\author{C.~Armand}\affiliation{Laboratoire d'Annecy de Physique des Particules, Univ. Grenoble Alpes, Univ. Savoie Mont Blanc, CNRS, LAPP, 74000 Annecy, France}
\author[0000-0001-5067-2620]{T.~Armstrong}\affiliation{University of Oxford, Department of Physics, Denys Wilkinson Building, Keble Road, Oxford OX1 3RH, UK}
\author[0000-0002-2153-1818]{H.~Ashkar}\affiliation{IRFU, CEA, Universit\'e Paris-Saclay, F-91191 Gif-sur-Yvette, France}
\author[0000-0002-9326-6400]{M.~Backes}\affiliation{University of Namibia, Department of Physics, Private Bag 13301, Windhoek 10005, Namibia}\affiliation{Centre for Space Research, North-West University, Potchefstroom 2520, South Africa}
\author{V.~Baghmanyan}\affiliation{Instytut Fizyki J\c{a}drowej PAN, ul. Radzikowskiego 152, 31-342 Krak{\'o}w, Poland}
% ...
\author[0000-0002-7130-9270
]{J.~N.S.~Shapopi}\affiliation{University of Namibia, Department of Physics, Private Bag 13301, Windhoek 10005, Namibia}
\author{K.~Shiningayamwe}\affiliation{University of Namibia, Department of Physics, Private Bag 13301, Windhoek 10005, Namibia}
\author{R.~Simoni}\affiliation{GRAPPA, Anton Pannekoek Institute for Astronomy, University of Amsterdam,  Science Park 904, 1098 XH Amsterdam, The Netherlands}
\author{A.~Sinha}\affiliation{Université de Paris, CNRS, Astroparticule et Cosmologie, F-75013 Paris, France}
\author{H.~Sol}\affiliation{Laboratoire Univers et Théories, Observatoire de Paris, Université PSL, CNRS, Université de Paris, 92190 Meudon, France}
\author{H.~Spackman}\affiliation{University of Oxford, Department of Physics, Denys Wilkinson Building, Keble Road, Oxford OX1 3RH, UK}
\author{A.~Specovius}\affiliation{Friedrich-Alexander-Universit\"at Erlangen-N\"urnberg, Erlangen Centre for Astroparticle Physics, Erwin-Rommel-Str. 1, D 91058 Erlangen, Germany}
\author[0000-0001-5516-1205]{S.~Spencer}\affiliation{University of Oxford, Department of Physics, Denys Wilkinson Building, Keble Road, Oxford OX1 3RH, UK}
\author{M.~Spir-Jacob}\affiliation{Université de Paris, CNRS, Astroparticule et Cosmologie, F-75013 Paris, France}
\author{{\L.}~Stawarz}\affiliation{Obserwatorium Astronomiczne, Uniwersytet Jagiello{\'n}ski, ul. Orla 171, 30-244 Krak{\'o}w, Poland}
\author{R.~Steenkamp}\affiliation{University of Namibia, Department of Physics, Private Bag 13301, Windhoek 10005, Namibia}
\author{C.~Stegmann}\affiliation{Institut f\"ur Physik und Astronomie, Universit\"at Potsdam,  Karl-Liebknecht-Strasse 24/25, D 14476 Potsdam, Germany}\affiliation{DESY, D-15738 Zeuthen, Germany}
\author[0000-0002-2865-8563]{S.~Steinmassl}\affiliation{Max-Planck-Institut f\"ur Kernphysik, P.O. Box 103980, D 69029 Heidelberg, Germany}
\author{C.~Steppa}\affiliation{Institut f\"ur Physik und Astronomie, Universit\"at Potsdam,  Karl-Liebknecht-Strasse 24/25, D 14476 Potsdam, Germany}
\author{L.~Sun}\affiliation{GRAPPA, Anton Pannekoek Institute for Astronomy, University of Amsterdam,  Science Park 904, 1098 XH Amsterdam, The Netherlands}
\author{T.~Takahashi}\affiliation{Kavli Institute for the Physics and Mathematics of the Universe (WPI), The University of Tokyo Institutes for Advanced Study (UTIAS), The University of Tokyo, 5-1-5 Kashiwa-no-Ha, Kashiwa, Chiba, 277-8583, Japan}
\author{T.~Tanaka}\affiliation{Kavli Institute for the Physics and Mathematics of the Universe (WPI), The University of Tokyo Institutes for Advanced Study (UTIAS), The University of Tokyo, 5-1-5 Kashiwa-no-Ha, Kashiwa, Chiba, 277-8583, Japan}
\author{T.~Tavernier}\affiliation{IRFU, CEA, Universit\'e Paris-Saclay, F-91191 Gif-sur-Yvette, France}
\author[0000-0001-9473-4758]{A.M.~Taylor}\affiliation{DESY, D-15738 Zeuthen, Germany}
\author[0000-0002-8219-4667]{R.~Terrier}\affiliation{Université de Paris, CNRS, Astroparticule et Cosmologie, F-75013 Paris, France}
\author{C.~Thorpe~Morgan}\affiliation{Institut f\"ur Astronomie und Astrophysik, Universit\"at T\"ubingen, Sand 1, D 72076 T\"ubingen, Germany}
\author{J.~H.E.~Thiersen}\affiliation{Centre for Space Research, North-West University, Potchefstroom 2520, South Africa}
\author{M.~Tluczykont}\affiliation{Universit\"at Hamburg, Institut f\"ur Experimentalphysik, Luruper Chaussee 149, D 22761 Hamburg, Germany}
\author{L.~Tomankova}\affiliation{Friedrich-Alexander-Universit\"at Erlangen-N\"urnberg, Erlangen Centre for Astroparticle Physics, Erwin-Rommel-Str. 1, D 91058 Erlangen, Germany}
\author{M.~Tsirou}\affiliation{Laboratoire Univers et Particules de Montpellier, Universit\'e Montpellier, CNRS/IN2P3,  CC 72, Place Eug\`ene Bataillon, F-34095 Montpellier Cedex 5, France}
\author{M.~Tsuji}\affiliation{Department of Physics, Rikkyo University, 3-34-1 Nishi-Ikebukuro, Toshima-ku, Tokyo 113-0033, 171-8501, Japan}
\author{R.~Tuffs}\affiliation{Max-Planck-Institut f\"ur Kernphysik, P.O. Box 103980, D 69029 Heidelberg, Germany}
\author{Y.~Uchiyama}\affiliation{Department of Physics, Rikkyo University, 3-34-1 Nishi-Ikebukuro, Toshima-ku, Tokyo 113-0033, 171-8501, Japan}
\author{D.J.~van~der~Walt}\affiliation{Centre for Space Research, North-West University, Potchefstroom 2520, South Africa}
\author[0000-0001-9669-645X]{C.~van~Eldik}\affiliation{Friedrich-Alexander-Universit\"at Erlangen-N\"urnberg, Erlangen Centre for Astroparticle Physics, Erwin-Rommel-Str. 1, D 91058 Erlangen, Germany}
\author{C.~van~Rensburg}\affiliation{University of Namibia, Department of Physics, Private Bag 13301, Windhoek 10005, Namibia}
\author{B.~van~Soelen}\affiliation{Department of Physics, University of the Free State,  PO Box 339, Bloemfontein 9300, South Africa}
\author{G.~Vasileiadis}\affiliation{Laboratoire Univers et Particules de Montpellier, Universit\'e Montpellier, CNRS/IN2P3,  CC 72, Place Eug\`ene Bataillon, F-34095 Montpellier Cedex 5, France}
\author{J.~Veh}\affiliation{Friedrich-Alexander-Universit\"at Erlangen-N\"urnberg, Erlangen Centre for Astroparticle Physics, Erwin-Rommel-Str. 1, D 91058 Erlangen, Germany}
\author{C.~Venter}\affiliation{Centre for Space Research, North-West University, Potchefstroom 2520, South Africa}
\author{P.~Vincent}\affiliation{Sorbonne Universit\'e, Universit\'e Paris Diderot, Sorbonne Paris Cit\'e, CNRS/IN2P3, Laboratoire de Physique Nucl\'eaire et de Hautes Energies, LPNHE, 4 Place Jussieu, F-75252 Paris, France}
\author{A.~Viana}\affiliation{Now at Instituto de Física de S\~ao Carlos, Universidade de S\~ao Paulo, Av. Trabalhador S\~ao-carlense, 400 - CEP 13566-590, S\~ao Carlos, SP, Brasil}
\author{J.~Vink}\affiliation{GRAPPA, Anton Pannekoek Institute for Astronomy, University of Amsterdam,  Science Park 904, 1098 XH Amsterdam, The Netherlands}
\author[0000-0003-2386-8067]{H.J.~V\"olk}\affiliation{Max-Planck-Institut f\"ur Kernphysik, P.O. Box 103980, D 69029 Heidelberg, Germany}
\author[0000-0002-7474-6062]{S.J.~Wagner}\affiliation{Landessternwarte, Universit\"at Heidelberg, K\"onigstuhl, D 69117 Heidelberg, Germany}
\author{F.~Werner}\affiliation{Max-Planck-Institut f\"ur Kernphysik, P.O. Box 103980, D 69029 Heidelberg, Germany}
\author{R.~White}\affiliation{Max-Planck-Institut f\"ur Kernphysik, P.O. Box 103980, D 69029 Heidelberg, Germany}
\author[0000-0003-4472-7204]{A.~Wierzcholska}\affiliation{Instytut Fizyki J\c{a}drowej PAN, ul. Radzikowskiego 152, 31-342 Krak{\'o}w, Poland}\affiliation{Landessternwarte, Universit\"at Heidelberg, K\"onigstuhl, D 69117 Heidelberg, Germany}
\author{Yu~Wun~Wong}\affiliation{Friedrich-Alexander-Universit\"at Erlangen-N\"urnberg, Erlangen Centre for Astroparticle Physics, Erwin-Rommel-Str. 1, D 91058 Erlangen, Germany}
\author{H.~Yassin}\affiliation{Centre for Space Research, North-West University, Potchefstroom 2520, South Africa}
\author{A.~Yusafzai}\affiliation{Friedrich-Alexander-Universit\"at Erlangen-N\"urnberg, Erlangen Centre for Astroparticle Physics, Erwin-Rommel-Str. 1, D 91058 Erlangen, Germany}
\author[0000-0001-5801-3945]{M.~Zacharias}\affiliation{Centre for Space Research, North-West University, Potchefstroom 2520, South Africa}\affiliation{Laboratoire Univers et Théories, Observatoire de Paris, Université PSL, CNRS, Université de Paris, 92190 Meudon, France}
\author{R.~Zanin}\affiliation{Max-Planck-Institut f\"ur Kernphysik, P.O. Box 103980, D 69029 Heidelberg, Germany}
\author[0000-0002-2876-6433]{D.~Zargaryan}\affiliation{Dublin Institute for Advanced Studies, 31 Fitzwilliam Place, Dublin 2, Ireland}\affiliation{High Energy Astrophysics Laboratory, RAU,  123 Hovsep Emin St  Yerevan 0051, Armenia}
\author{A.A.~Zdziarski}\affiliation{Nicolaus Copernicus Astronomical Center, Polish Academy of Sciences, ul. Bartycka 18, 00-716 Warsaw, Poland}
\author{A.~Zech}\affiliation{Laboratoire Univers et Théories, Observatoire de Paris, Université PSL, CNRS, Université de Paris, 92190 Meudon, France}
\author[0000-0002-6468-8292]{S.J.~Zhu}\affiliation{DESY, D-15738 Zeuthen, Germany}
\author{A.~Zmija}\affiliation{Friedrich-Alexander-Universit\"at Erlangen-N\"urnberg, Erlangen Centre for Astroparticle Physics, Erwin-Rommel-Str. 1, D 91058 Erlangen, Germany}
\author{J.~Zorn}\affiliation{Max-Planck-Institut f\"ur Kernphysik, P.O. Box 103980, D 69029 Heidelberg, Germany}
\author[0000-0002-5333-2004]{S.~Zouari}\affiliation{Université de Paris, CNRS, Astroparticule et Cosmologie, F-75013 Paris, France}
\author{N.~\.Zywucka}\affiliation{Centre for Space Research, North-West University, Potchefstroom 2520, South Africa}
\collaboration{275}{(H.E.S.S. Collaboration)}

\correspondingauthor{D.~Glawion, D.~Malyshev, A.~Montanari, E.~Moulin}
\email{contact.hess@hess-experiment.eu}

\date{\today}

\begin{abstract}
Cosmological $N$-body simulations show that Milky Way-sized galaxies harbor a population of unmerged dark matter subhalos. These subhalos could shine in gamma-rays
and be eventually detected in gamma-ray surveys as unidentified 
sources. 
We performed a thorough selection among unidentified \textit{Fermi}-LAT Objects (UFOs) to identify them as
possible TeV-scale dark matter subhalo candidates.
We search for very-high-energy (E $\gtrsim$ 100~GeV) gamma-ray emissions using H.E.S.S. observations towards four selected UFOs.
Since no significant very-high-energy gamma-ray emission is detected in any dataset of the four observed UFOs nor in the combined UFO dataset, strong constraints are derived on the product of the velocity-weighted annihilation cross section $\langle \sigma v \rangle$ by the $J$-factor for the
dark matter models. The 95\% C.L. observed upper limits derived from combined H.E.S.S. observations reach $\langle \sigma v \rangle J$ values of 
3.7$\times$10$^{-5}$ and 8.1$\times$10$^{-6}$ GeV$^2$cm$^{-2}$s$^{-1}$ in the $W^+W^-$ and $\tau^+\tau^-$ channels, respectively, for a 1~TeV dark matter mass.
Focusing on thermal WIMPs, the H.E.S.S. constraints restrict the $J$-factors to lie in the range 6.1$\times$10$^{19}$ - 2.0$\times$10$^{21}$ GeV$^2$cm$^{-5}$, and the masses to lie between 0.2 and 6 TeV in the  $W^+W^-$ channel. For the  $\tau^+\tau^-$ channel, the $J$-factors lie in the range 7.0$\times$10$^{19}$ - 7.1$\times$10$^{20}$ GeV$^2$cm$^{-5}$ and the masses lie between 0.2 and 0.5 TeV.
Assuming model-dependent predictions from cosmological N-body simulations on the $J$-factor distribution for Milky Way-sized galaxies, the dark matter models with masses greater than 0.3 TeV for the UFO emissions  
can be ruled out at high confidence level.
\end{abstract}


\section{Introduction}
\label{sec:intro}
The presence of dark matter (DM) is suggested by a wealth of astrophysical and cosmological measurements, however its underlying nature is yet unknown. Among the most promising candidates are weakly interacting massive particles (WIMPs): 
particles thermally-produced in the early universe with mass and coupling strength at the electroweak scale predict a present relic density~\citep{Steigman:2012nb} consistent with that observed today~\citep{Adam:2015rua}. WIMP self-annihilation would produce Standard Model particles including gamma rays which for a long time, have been 
recognized as a prime messenger to indirectly detect DM annihilation or decay.
Gamma rays are not deflected by magnetic fields and therefore point back to their sources. Among the most promising DM targets observed by ground-based imaging atmospheric Cherenkov telescopes (IACTs) such as H.E.S.S. are the Galactic Centre~\citep{Abdallah:2016ygi,Abdallah:2018qtu} and nearby dwarf
galaxies~\citep{Aharonian:2007km,Abramowski:2010aa,Abramowski:2014tra,Abdalla:2018mve}.

Other compelling and complementary DM targets for IACTs are DM subhalos populating galactic halo~\citep[see, e.g.,][]{Kamionkowski:2010mi}. The observed universe today is believed to have formed hierarchically with the smallest structures first. DM particles first collapse into gravitationally-bound systems that later merge to form the first subhalos, which subsequently form more massive ones. The merging history leads to DM halos massive enough to retain gas and trigger star formation and give rise to the galaxies we observe today. However, most of the subhalos remain completely dark. Assuming that DM is made of WIMPs, 
they could shine in gamma rays. The annihilation process in subhalos could be frequent enough to be detectable by IACTs provided that the WIMPs are sufficiently massive. If DM subhalos are made of WIMPs, 10$^{-4}$ to 10$^{10}$ M$_{\odot}$ mass subhalos are expected to lie in DM halos of Milky Way-sized galaxies~\citep{Diemand:2008in,Springel:2008cc}, with the most massive of them ($\gtrsim 10^8$ M$_{\odot}$) hosting dwarf galaxies. 

DM subhalos are predicted to be compact and concentrated, and are not expected to harbour conventional astrophysical high-energy emitters. Provided they are close and/or massive enough, DM annihilations in these objects could produce 
gamma-ray fluxes detectable with satellite and ground-based experiments such as  IACTs~\citep{CalcaneoRoldan:2000yt,Tasitsiomi:2002vh,Stoehr:2003hf,Koushiappas:2003bn}. However, their actual location in the galaxy is not known. 
Their search can be performed using all-sky gamma-ray observations~\citep[see, for instance,][]{Diemand:2006ik} such as with the Large Area Telescope (LAT) instrument onboard the Fermi satellite~\citep[see, for instance,][]{Berlin:2013dva} 
or wide-field surveys carried out with IACTs~\citep[see, for instance,][]{Aharonian:2008wt,Brun:2010ci}.

All-sky \flat observations revealed a population of sources that lack associated signals from observations at other wavelengths~\citep{TheFermi-LAT:2017pvy,Fermi-LAT:2019yla}. These sources are therefore classified as unidentified Fermi objects (UFOs). 
If the DM particle mass lies below 100 GeV, some UFOs detected by \flat could be potentially described by DM models~\citep{Belikov:2011pu,Zechlin:2011kk,Bertoni:2015mla,Bertoni:2016hoh,Calore:2016ogv,Coronado-Blazquez:2019puc}
Identifying some of the UFOs as DM subhalos require however higher ($\gtrsim 100$~GeV) DM masses given their hard gamma-ray spectra in the few-ten-to-hundred GeV energy range.
Such objects are therefore excellent targets for IACTs to perform searches for TeV DM subhalos. In 2018 and 2019, the H.E.S.S. collaboration carried out an observational campaign for a selection of the most promising UFOs in order to probe their potential TeV-mass DM-induced emission. 

The paper is organized as follows. Section~\ref{sec:darkmatter} presents the expected DM-induced gamma-ray signals from Galactic subhalos. Section~\ref{sec:ufos} describes the selection procedure of UFOs as DM subhalo candidates relevant for H.E.S.S. observations along with the \flat data analysis of the selected UFOs as DM subhalo candidates. In Sec.~\ref{sec:analysis}, the H.E.S.S. observations and data analysis of the selected UFOs are presented. The constraints from H.E.S.S. observations on DM-induced emission models for the UFOs are derived in Sec.~\ref{sec:results}. Section~\ref{sec:discussion} and Sec.~\ref{sec:summary} are devoted to the discussion of the results obtained in this work and to a general summary, respectively. 

\section{Dark matter annihilation signals}
\label{sec:darkmatter}
\subsection{Expected gamma-ray flux from dark matter annihilation}
The energy-differential gamma-ray flux expected from the self-annihilation of Majorana DM particles of mass $m_{\rm DM}$ can be expressed as
\begin{equation}
\label{eq:dmflux}
\frac{{\rm d} \Phi_\gamma}{{\rm d} E_\gamma} (E_\gamma,\Delta\Omega)=
\frac {\langle \sigma v \rangle}{8\pi m_{\rm DM}^2}\sum_f  \text{BR}_f \frac{{\rm d} N^f}{{\rm d}E_\gamma} \, J(\Delta\Omega) \ ,
\quad {\rm with} \quad  J(\Delta\Omega) =  \int_{\Delta\Omega} \int_{\rm los}
\rho^2(s(r,\theta)) ds\, d\Omega \, .
\end{equation}
$\langle \sigma v \rangle$ is the thermally-averaged  velocity-weighted annihilation cross section, and $\sum_f  \text{BR}_f  dN^f/dE_{\gamma}$ is the sum of the annihilation spectra ${\rm d}N^f/{\rm d}E_{\gamma}$ per annihilation in the final states $f$ with associated branching ratios ${\rm BR}_{f}$. The expected DM annihilation signal consists of a continuum spectrum of gamma rays extending up to the DM mass, and possibly a line-like feature close to the DM mass. The former contribution arises from the hadronization and/or decay of quarks, heavy leptons, and gauge bosons involved in the annihilation process.
The latter comes from the direct annihilation into $\gamma X$ with X= $\gamma$ , h, Z or a non-Standard Model neutral particle, providing a  spectral line at an energy $E_{\gamma} =  m_{\rm DM}[1-(m_X/2m_{\rm DM})^2]$. 
When the DM particles self-annihilate into charged particles, gamma rays are produced via processes involving virtual internal bremsstrahlung and final state radiation. These processes provide an additional bump-like feature that peaks at an energy close to the DM mass.

The term $J(\Delta\Omega)$, hereafter referred to 
the $J$-factor, corresponds to the integration of the square of the DM density over the line-of-sight (los) $s$ and solid angle $\Delta\Omega$. As opposed to objects with measured stellar dynamics like dwarf galaxies, 
UFOs have unknown distances to Earth and their $J$-factors cannot be derived from stellar kinematics.

\subsection{Expected subhalo $J$-factor distribution in the Milky Way}
\label{sec:subproperties}
Cosmological $N$-body simulations (see, for instance, Refs.~\citep{Diemand:2008in,Springel:2008cc}) predict Milky Way (MW)-sized galaxy halos to harbor today unmerged DM substructures called galactic subhalos. These simulations make robust predictions on the slope and normalization of the subhalo mass function defined as ${\rm d}\ln N /{\rm d} \ln M \propto M^{-\alpha_{\rm m}}$, with a slope $\alpha_{\rm m}$ $\simeq$1.9 for MW-like galaxies (see, for instance, Refs.~\citep{Diemand:2008in,Springel:2008by,Gao:2012tc,Fiacconi:2016jvv}). 
Galactic subhalos are not expected to host conventional high-energy astrophysical sources and a large number density of subhalos in MW-like galaxies with high DM concentrations is expected. 
Only closeby and massive enough subhalos are expected to provide detectable gamma-ray signals. 

The cosmological simulations provide the abundance of the resolved subhalos, their radial distribution and 
structural properties such as their mass and concentration, and suggest a DM distribution in the subhalos following a cuspy density profile,
that can be well described by NFW~\citep{Navarro:1996gj} or 
Einasto~\citep{springel2008}
parametrizations. The limited spatial resolution of the current simulations can only probe the slope of the radial DM density distribution in a subhalo for the most massive subhalos. Assuming a parametrization of the DM density distribution in subhalos, the distribution of $J$-factors of the galactic subhalo population, $dN/dJ$, can be derived. The cumulative $J$-factor distribution, $N(J)\equiv N(\geq J)$, is defined as number of subhalos with a $J$-factor higher than or equal to specified.

In order to derive the $J$-factor distribution 
of DM subhalos in the MW, the CLUMPY code v3.0.0~\citep{clumpy,clumpy_v2,clumpy_v3} is used.  
1000 simulations of a MW-like galaxy are performed with a smooth NFW~\citep{Navarro:1996gj} DM main halo profile with the parameters corresponding to the best-fit NFW parameters presented in a recent study of DM distribution in the Milky Way~\citep{cautun20}. 
For each simulation, the subhalo parameters were chosen similar to the ones used in  Ref.~\citep{hutten16} for the ``HIGH'' model. The power-law slope of the subhalo mass function is chosen to be $\alpha_{\rm m} = 1.9$~\citep{Diemand:2008in}; the number of objects between $10^8$ and $10^{10}M_{\odot}$ is taken as $N_{\rm calib}$ = 300 following \cite{Springel:2008cc}; and the subhalo mass-concentration relation following the distance-dependent prescription of~\cite{moline17}.
From each simulation the Galactic coordinates of all  subhalos and their $J$-factors 
integrated in circular regions with 0.1$^\circ$ radius around the centres of gravity of the subhalos are derived.

The cumulative $J$-factor distribution $N(\geq J)$ is shown in the upper panel of Fig.~\ref{fig:luminosty_function} for subhalos located at Galactic latitudes $|b| > 5^\circ$. 
The solid red curve shows the averaged distribution computed from all the realizations and the shaded region shows the formal 1$\sigma$ statistical dispersion calculated over all simulated MW-like galaxies. In the lower panel of Fig.~\ref{fig:luminosty_function}, the red-dashed curve illustrates the probability to find in any simulation at least one subhalo with a $J$-factor higher than specified. The blue-dotted curve corresponds to the probability to find three or more subhalos. The horizontal black-dashed line gives the 5\% probability: 
\begin{figure*}[hb!]
\begin{center}
\includegraphics[width=0.5\linewidth]{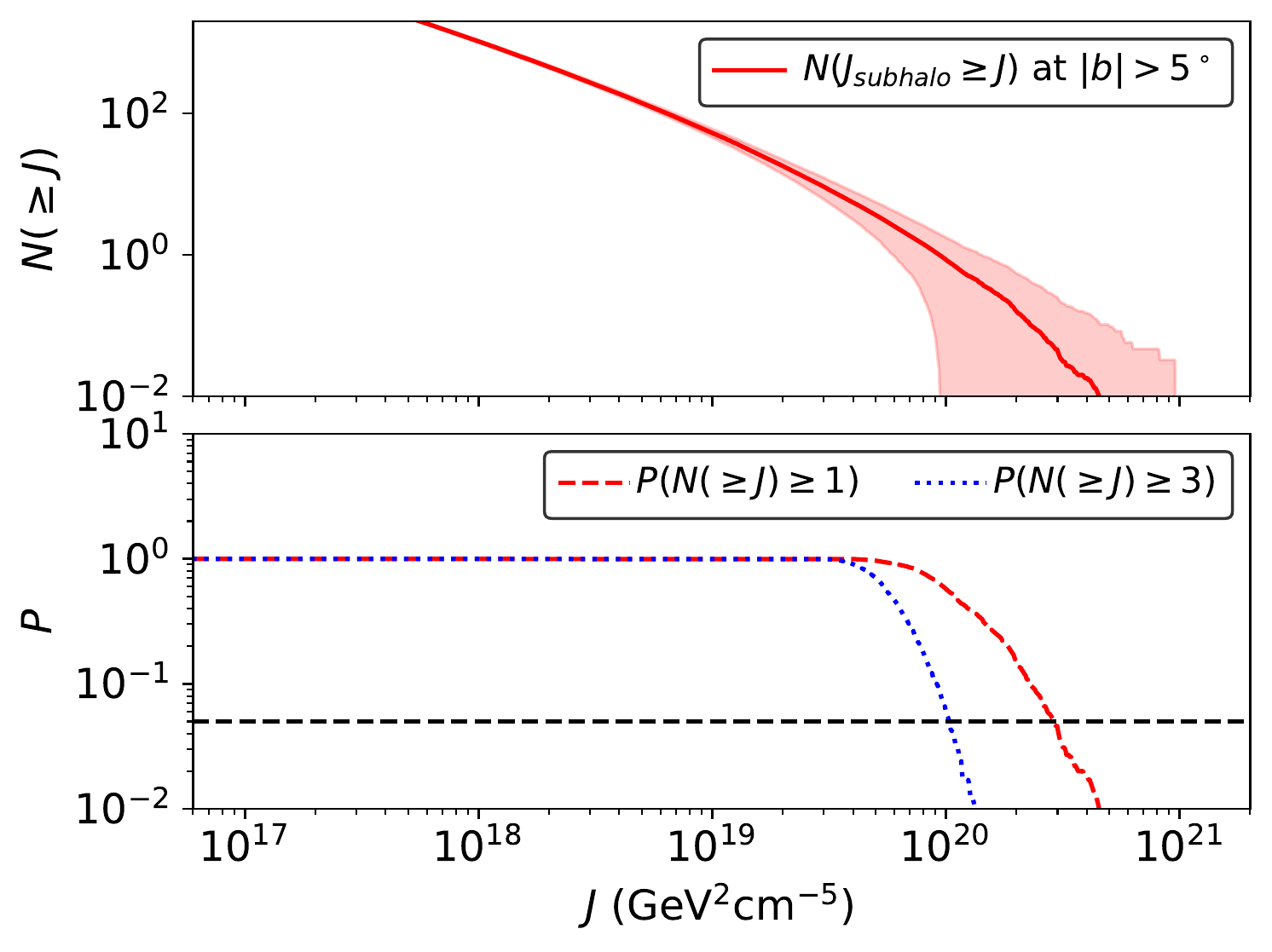}
\caption{{\it Top panel:} Cumulative $J$-factor distribution, $N(\ge J)$, of a Milky Way-like subhalo population. The number of subhalos with a $J$-factor exceeding a given value is plotted (red solid curve).
The red-shaded band corresponds to the 1$\sigma$ statistical uncertainty obtained from 1000 simulations of the subhalo population for MW-like galaxies. 
{\it Bottom panel:}  
Probability $P$ to find at least one subhalo with a $J$-factor higher than specified (red dashed line). The blue dotted line presents the same probability for at least three suhhalos. The horizontal black dashed line shows the 5\%  probability. See text for more details.}
\label{fig:luminosty_function}
\end{center}
\end{figure*}
the probability on average to find one or more subhalos with $J$-factor $J \ge 3\times 10^{20}$~GeV$^2$cm$^{-5}$ (three or more subhalos with $J$-factors $J \ge 1\times 10^{20}$~GeV$^2$cm$^{-5}$) is only about 5\%.

\section{\flat unidentified sources as dark matter subhalo candidates}
\label{sec:ufos}
A hypothetical gamma-ray emission from a DM subhalo would show-up 
in the all-sky gamma-ray surveys~\citep{Kamionkowski:2010mi} as unidentified objects, {\it i.e.}, detected with the Fermi satellite but without counterparts at any other wavelengths. A smoking-gun signature for DM detection is a very distinct energy cut-off close to the DM particle mass, assuming the two-body annihilation process taking place almost at rest. 
For sufficiently-large DM particle masses, {\it i.e.}, above a few hundred GeV,
the energy cut-off could be too high in energy to be measurable by the LAT instrument onboard the Fermi satellite within a reasonable observation time. The combination of  \flat and IACT observations is therefore mandatory for DM subhalo searches with unidentified sources detected by the \flat.

\subsection{Selection procedure for H.E.S.S. observations}
\label{sec:selection}
In order to determine the best candidates for H.E.S.S. observations among the unidentified \flat sources, a thorough selection has been performed in the Third Catalog of High-Energy \flat Sources (3FHL)~\citep{TheFermi-LAT:2017pvy}
which comprises pointlike sources detected above 10 GeV.
The source selection requires 
the unidentified sources to be steady, {\it i.e.}, to not show flux variability over time according to 3FHL catalogue\footnote{While the criterium on the variability provides steady candidates as expected for DM sources, \flat photon properties at the highest energies have been checked. None of them could be attributed to flaring of nearby \flat sources.},
exhibiting a hard power-law spectral index ($\Gamma < 2$), as expected for DM-induced signals for DM masses above 100 GeV with no obvious conventional counterpart at other wavelengths. 
The multi-wavelength (MWL) search for possible counterparts is based on the Fermi-LAT source coordinates : for each source, counterparts are searched in catalogs of MWL facilities (\textit{XMM-Newton}, \textit{ROSAT}, \textit{SUZAKU}, \textit{CGRO}, \textit{Chandra}, \textit{Swift}, \textit{WMAP}, \textit{RXTE}, \textit{Nustar}, \textit{SDSS}, \textit{Planck}, \textit{WISE}, \textit{HST}) assuming a searched radius around the source determined by the position uncertainty derived in the 3FHL catalogue.
The unidentified \flat sources are usually quite faint and only the ones sufficiently far away from the Galactic plane are considered in order to avoid challenging background modelling connected to Galactic plane diffuse emission in the \flat energy range. The selected sources do not lie in a complex astrophysical environment, {\it i.e.}, they are relatively isolated with no high-energy gamma-ray emission within about one degree\footnote{The closest 3FHL source for 3FHL J1915.2-1323 is at 0.8$^\circ$ while for the other UFOs, the closest source is at distance higher than 1.7$^\circ$).}. 
In addition, a maximum zenith angle of $45^\circ$ for H.E.S.S. observations is required in order to obtain low energy thresholds.
The selection criteria were applied on the 3FHL source catalog and are summarized in Tab.~\ref{tab:table1}.
The small number of suitable DM subhalo candidates obtained by a straightforward selection confirms that the observation of a selection of UFOs is a viable DM search strategy for targeted observations performed by IACTs. 
Only four UFOs were eventually observed due to H.E.S.S. observational program scheduling constraints. The characteristics of the UFOs selected for observations with the H.E.S.S. telescopes are summarized in Tab.~\ref{tab:table2}.
\begin{table}[hb!]
\centering
{\scriptsize
\begin{tabular}{lc}     
\hline
\hline
Criteria  & Numbers of sources \\\hline
Without association &  178 \\
Far enough from the Galactic plane, cut in Galactic latitude of $|b|>5^\circ$  & 126\\
Non-variable, cut in variability 
index (No. of Bayesian blocks in var. analysis) equal to 1  & 125\\
Maximum zenith angle at H.E.S.S. site of $45^\circ$ & 83\\
Follow a simple power law with significance for curvature $<3\sigma$    & 83\\
Hard spectrum, cut in spectral index below 2 & 18\\
No MWL counterparts & 6\\
\hline
\hline
\end{tabular}
}
\caption{\label{tab:table1} Criteria applied to 3FHL catalog for the selection of DM subhalo candidates. For the multi-wavelength (MWL) counterpart search, individual search radii were used ($\sim2-4$ arcmin) based on the uncertainty of the  Fermi position quoted in the 3FHL. The following list of MWL facilities was checked: \textit{XMM-Newton}, \textit{ROSAT}, \textit{SUZAKU}, \textit{CGRO}, \textit{Chandra}, \textit{Swift}, \textit{WMAP}, \textit{RXTE}, \textit{Nustar}, \textit{SDSS}, \textit{Planck}, \textit{WISE}, \textit{HST}.}
\end{table}

\begin{table}[htb!]
\centering
{\scriptsize
\begin{tabular}{l | c | c | c | c | c | c| c |c| c}
\hline
\hline
Name & RA & Dec. & TS for & Position  & Pivot & Spectral energy distribution & Power-law  & $\Delta\chi^2$ & $E_{\rm cut}$\\
    &   &  & $E\geq10$~GeV  & uncertainty & energy & at pivot energy  &  index & & \\
          & [degrees] & [degrees] &  &[arcmin] & [GeV]  & [$10^{-13}$ TeV\,cm$^{-2}$s$^{-1}$] & & &[GeV]\\
\hline
3FHL J0929.2-4110 & 142.3345 &  -41.1833 & 36 & 2.4 & 0.39   & $0.12\pm 0.01$ & $1.37\pm 0.07$ & 0.15 & $>33$\\
3FHL J1915.2-1323$^\dagger$ & 288.8182 &  -13.3916 &23 & 3.0 & 62.8  &$2.1\pm 0.9$ & $1.5\pm 0.4$ & 0.05& $>35$\\
3FHL J2030.2-5037 & 307.5901 &  -50.6344 & 40 & 2.6 & 6.3 & $1.9\pm 0.3$& $1.85\pm 0.1$&0.40 & $>67$\\
3FHL J2104.5+2117\footnote{The spectral index in the 3FHL catalogue is 
1.8~\citep{TheFermi-LAT:2017pvy}.}$^,$\footnote{3FHL J2104.5.2117 was recently associated in the 4FGL catalogue~\citep{Fermi-LAT:2019yla} with an AGN with a probability of 0.4.
} & 316.1226 &  21.2831 & 58 & 2.2 & 1.56 & $5.3\pm 0.5$ & $2.22\pm 0.06$& 0.02& $>85$\\
\hline
\hline
\end{tabular}
}
\caption{\label{tab:table2} Properties of the selected UFOs together with their spectral parameters. The second and third columns provides the RA-Dec coordinates of the UFOs along with their test statistics (TS) values for energies above 10 GeV in the fourth column. The fifth column  gives their position uncertainty.  Pivot energy, spectral energy distribution at the pivot energy and best-fit power-law spectral index are given in the sixth, seventh, eighth columns, respectively. The ninth column provides the $\Delta \chi^2$ value between a pure power-law and a power law with exponential cut-off fit to the data. The last column gives the 95\% C.L. lower limit on a possible energy cut-off in the energy spectrum.
The 3FHL J1915.2-1323 source marked with $^\dagger$ is detected only above 10~GeV. For this source the spectral index, pivot energy, differential flux and $\Delta\chi^2$ value are given for this energy band. For the other sources these quantities are given for energies higher than 0.1~GeV.}
\end{table}

\subsection{\flat data analysis on the selected objects}
\label{sec:fermi_data_analysis}
\flat data selected for the analysis presented in this paper
were collected over a time span of more than 12 years (Aug. 2008 to Oct. 2020). The latest available \texttt{fermitools} v.~2.0.0  with P8R3\_V3 response functions (\texttt{CLEAN} photon class)\footnote{See \href{https://fermi.gsfc.nasa.gov/ssc/data/analysis/documentation/Cicerone/Cicerone_LAT_IRFs/c}{description of \flat response functions.}} are used.
The initial stage of the analysis aims at determining the energy spectra of UFO sources.
In order to extract the differential energy spectrum for each object the standard binned likelihood analysis of a 14$^\circ$-radius region around each of the considered objects 
is performed 
for a set of eight log-equal energy bins in the range $0.1 - 1000$~GeV. 
The spectral analysis is based on the fitting of a spatial and spectral model of the sky region around the source of interest to the data.  The model of the region includes: \textit{(i)}: all sources from the most recent 4FGL-DR2 catalogue~\citep{Fermi-LAT:2019yla} within the $14^\circ$-radius region around UFO position;  \textit{(ii):} components for isotropic and Galactic diffuse emissions given by the standard spatial and spectral templates \texttt{iso\_P8R3\_CLEAN\_V2\_v1.txt} and \texttt{gll\_iem\_v07.fits}, respectively. For the fitting procedure the spectral models of these sources are selected according to the 4FGL catalogue with all parameters except normalisation fixed to the catalogue values.
In addition 4FGL sources up to $10^\circ$ beyond the considered region of interest are included in the model, with all their parameters fixed to the catalogue values in order to reduce the bias connected to a possible presence of bright sources outside  the considered region and effects connected to the poor PSF of the LAT at low ($\sim 0.1$~GeV) energies.
The UFO spectra are modelled by a pure power-law function with the slope defined from a broad energy-range fit. 
\begin{figure}[!htbp]
\centering
\includegraphics[width=0.45\textwidth]{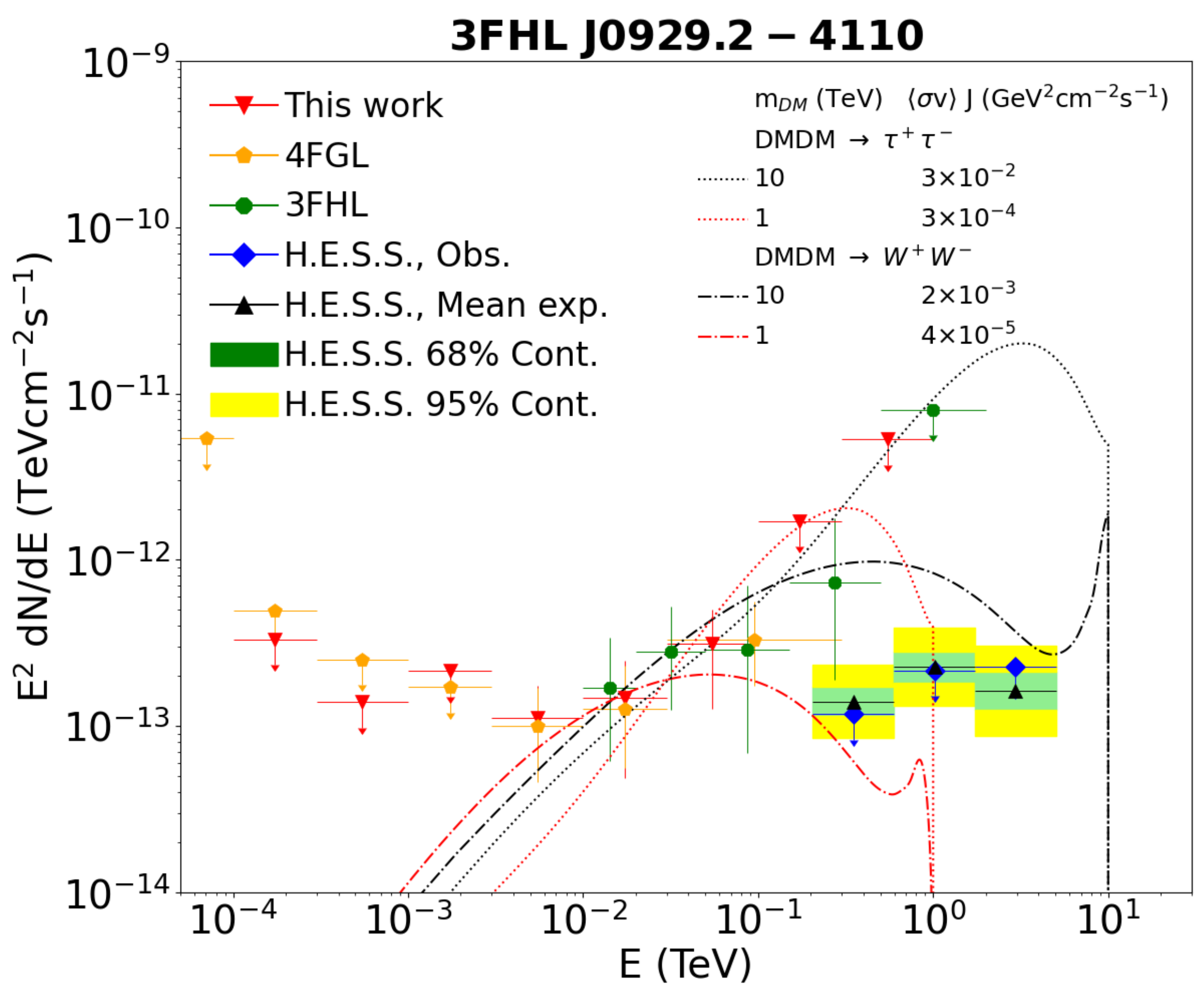}
\includegraphics[width=0.45\textwidth]{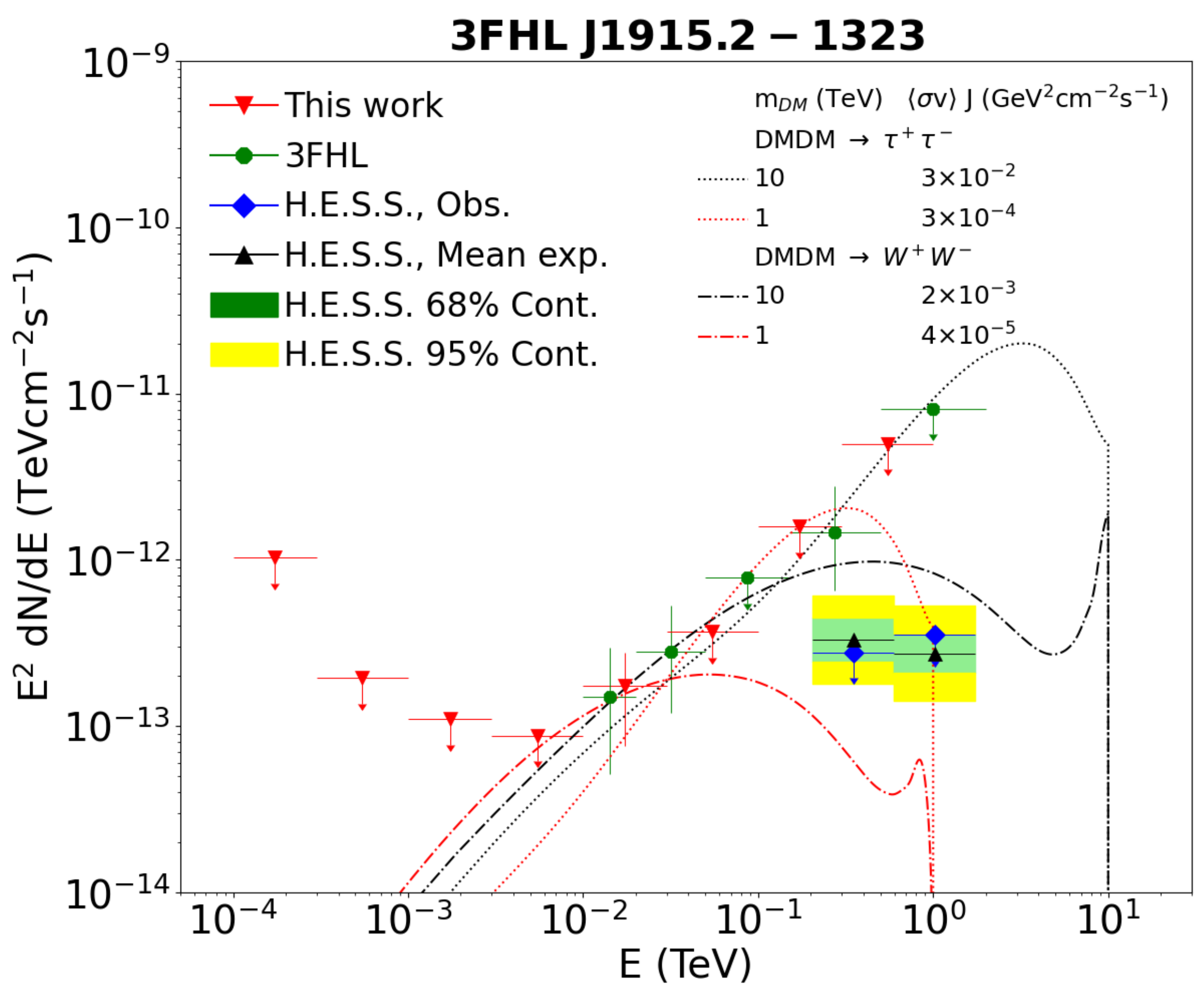}
\includegraphics[width=0.45\textwidth]{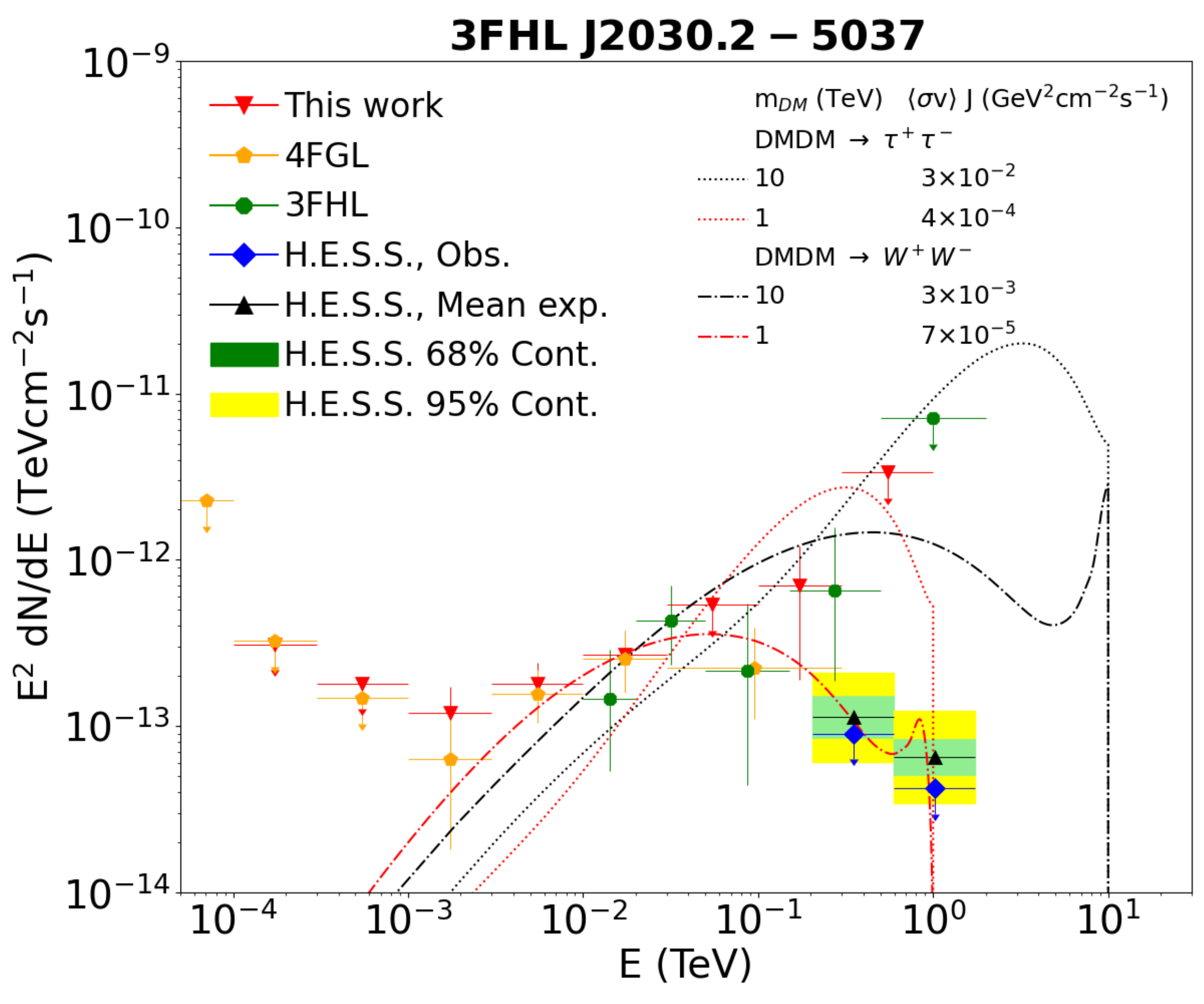}
\includegraphics[width=0.45\textwidth]{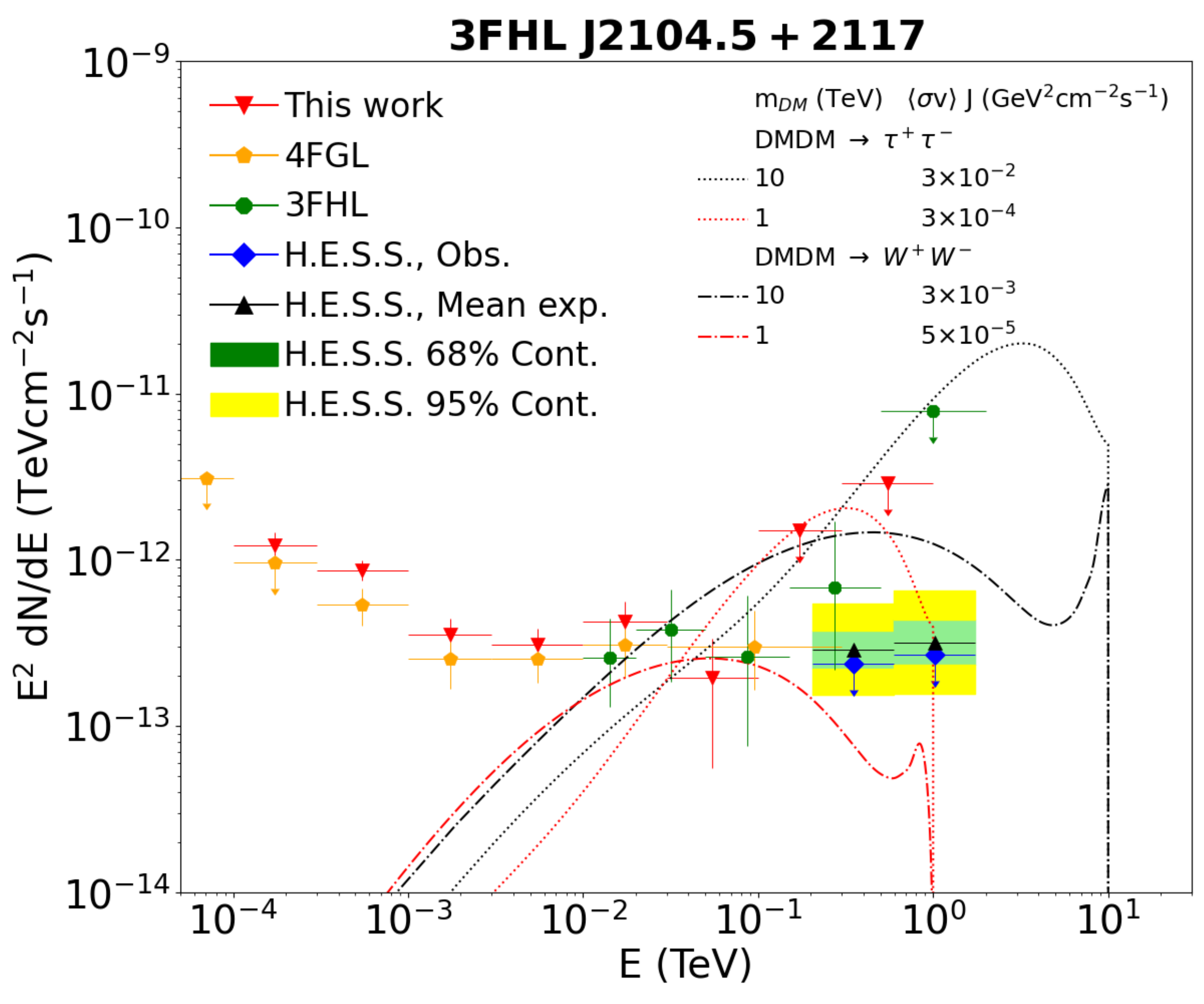}
\caption{Spectral energy distributions of the selected unidentified Fermi objects observed with \flat and H.E.S.S. for 3FHL J0929.2-4110 (top left), 3FHL J1915.2-1323 (top right), 3FHL J2030.2-5037 (bottom left), and 3FHL J2104.5+2117 (bottom right), respectively. The differential flux points computed in this work from the \flat dataset (red dots) and taken from the 4FGL (orange dots) and from the 3FHL (green dots) catalogues~\citep{TheFermi-LAT:2017pvy,Fermi-LAT:2019yla}, are shown with the vertical and horizontal error bars
corresponding to the 1$\sigma$ statistical errors and the bin size, respectively. Upper limit points (red, orange and green arrows) are given at
95\% C.L.. The observed flux upper limits from H.E.S.S. observations (blue arrows) are plotted at 95\% C.L., together with the 
mean expected flux upper limits (black) and the 1$\sigma$ (green) and 2$\sigma$ (yellow) containment bands.
Overlaid are theoretical DM-induced fluxes for 1 TeV and 10 TeV DM masses in the $W
^+W^-$ (dash-dotted lines) and $\tau^+\tau^-$ (dotted lines) annihilation channels, respectively.
}
\label{fig:sed}
\end{figure}
Following the recommendation of the \flat collaboration, our analysis is performed with energy dispersion handling enabled.
We additionally checked that the test-statistic maps of the considered regions do not show significant residuals between the data and the model, see Appendix~\ref{app:ts_maps} for more details. We conclude that the considered models well describe the UFO 
sources' regions.

Table~\ref{tab:table2} summarizes the analysis results for each UFO.
Given the available photon statistics in the \flat dataset of the selected objects, a power-law spectrum with a (super)-exponential energy cut-off in the TeV energy range, as expected from DM-induced emissions~\citep{Belikov:2013laa}, and a pure power-law emission cannot be significantly discriminated, see Tab.~\ref{tab:table2} for a corresponding change $\Delta\chi^2$ between these models. The last column of Tab.~\ref{tab:table2} summarises 95\% C.L. \flat lower limits  
on the  energy cutoff, defined as the energy at which $\Delta\chi^2$ changes by 2.71 between power-law and exponential energy cut-off power-law models.

\subsection{Dark matter models for the selected sources}
Figure~\ref{fig:sed} shows DM annihilation models 
together with \flat flux measurements. 
Model predictions for DM masses of 1 TeV and 10 TeV, respectively, are plotted separately for the $W^+W^-$ and $\tau^+\tau^-$ annihilation channels. Some DM models are able to qualitatively describe the observed gamma-ray flux from the selected UFOs.
For instance, the predictions shown for $m_{\rm DM}$ = 1 TeV describe well the \flat data for 3FHL J0929.2-4110 in the $W^+W^-$ and $\tau^+\tau^-$ annihilation channels. Instead,
For 3FHL J2030.2-5037, the predictions for $m_{\rm DM}$ = 1 
TeV in the $\tau^+\tau^-$ annihilation channel provide a poor description of the emission while a better one is obtained in the $W^+W^-$ one.

In order to assess quantitatively the above statements on viable DM-induced emission models, 
the spectrum of each UFO  is explicitly modelled with a DM-annihilation induced spectral template\footnote{Provided within fermitools as \texttt{DMFitFunction} based on Ref.~\citep{Jeltema_2008}}. For given  $m_{\rm DM}$ and annihilation channel, the model is characterised only by 
the overall normalisation of the spectra given by $\langle \sigma v\rangle J$. To identify the range of viable parameters for DM annihilation, a scan over a large range of $\langle \sigma v\rangle J$ is performed.

A test-statistic (TS) is defined as a difference between best-fit log-likelihood functions for models with no DM emission ($\mathcal{L}_0$, ``background only'' hypothesis) 
and the model ($\mathcal{L}$) which includes the UFO source described by the corresponding parameter $\langle \sigma v\rangle J$: $TS = -2\,\rm log (\mathcal{L}/\mathcal{L}_0)$~\citep{mattox96}\footnote{The $TS$ value for a source with $N$-parametric (spectral) model follows a $\chi^2$ distribution with $N$ degrees of freedom in the high statistic limit~\citep{wilks38}.}. 
 Negative TS values correspond to the detection of the source, {\it i.e.}, adding a source with a corresponding parameter \textit{improves} the fit in comparison to background-only hypothesis.

\begin{figure*}[ht!]
\centering
\includegraphics[width=0.48\textwidth]{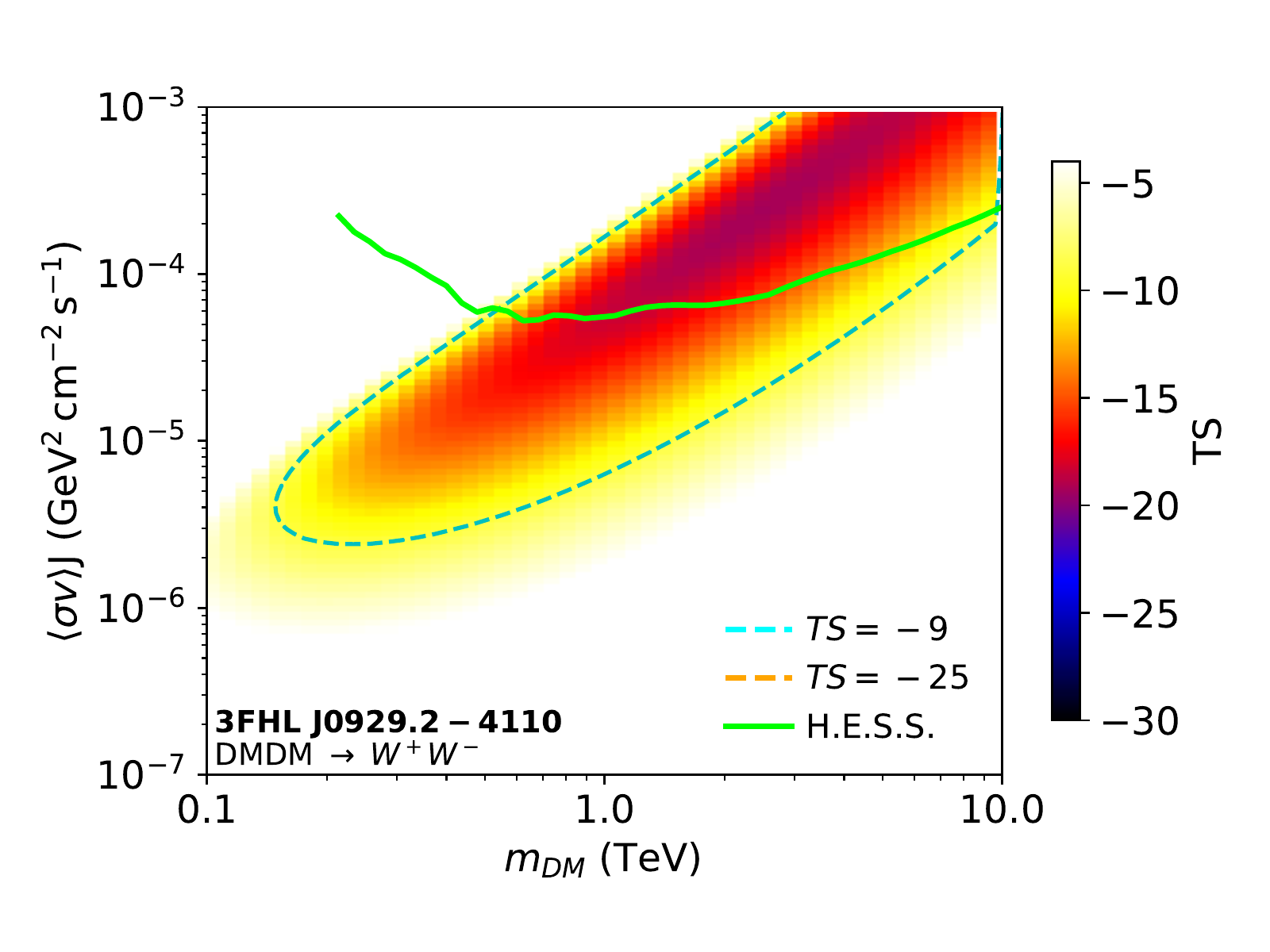}
\includegraphics[width=0.48\textwidth]{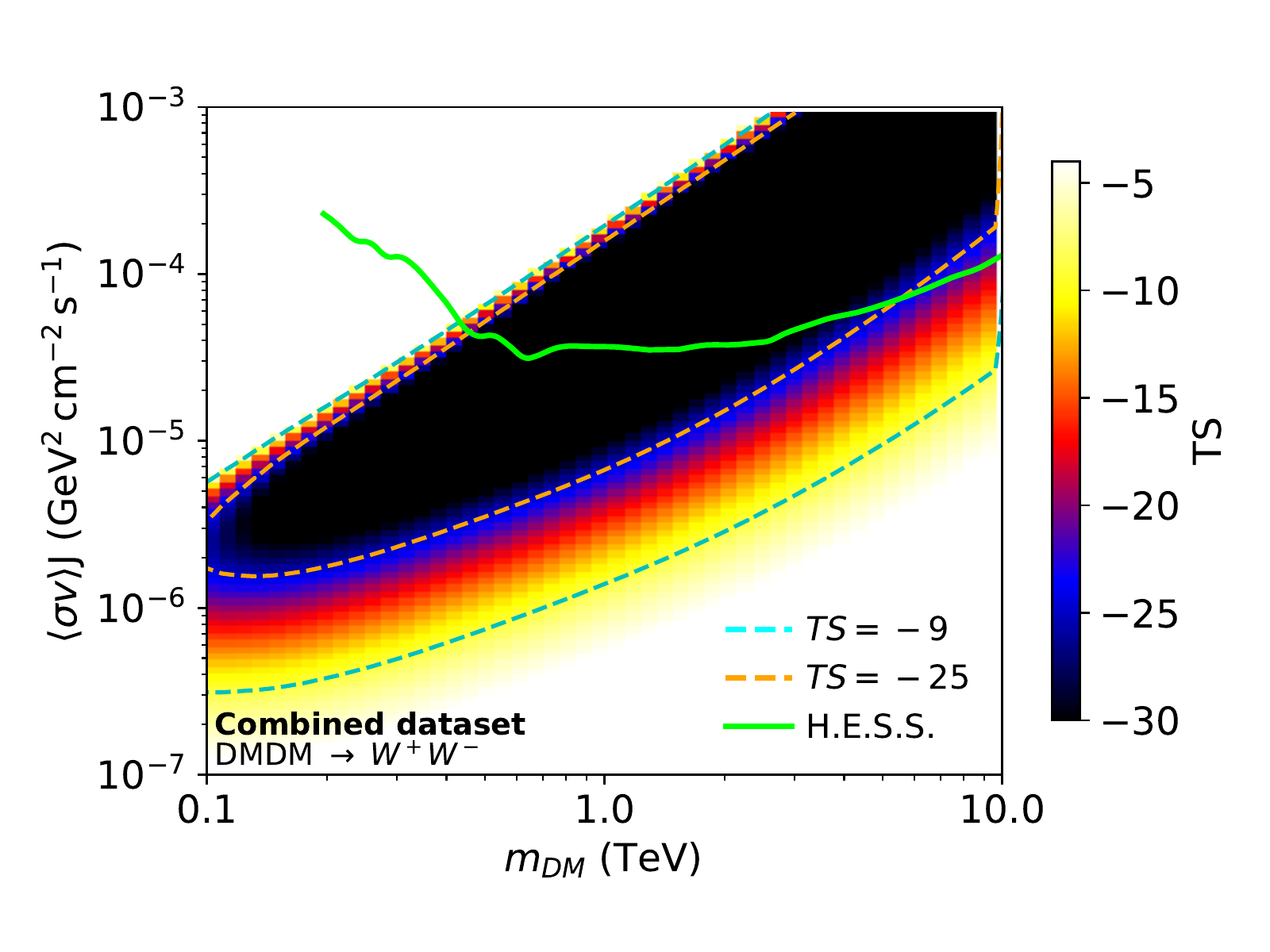}
\includegraphics[width=0.48\textwidth]{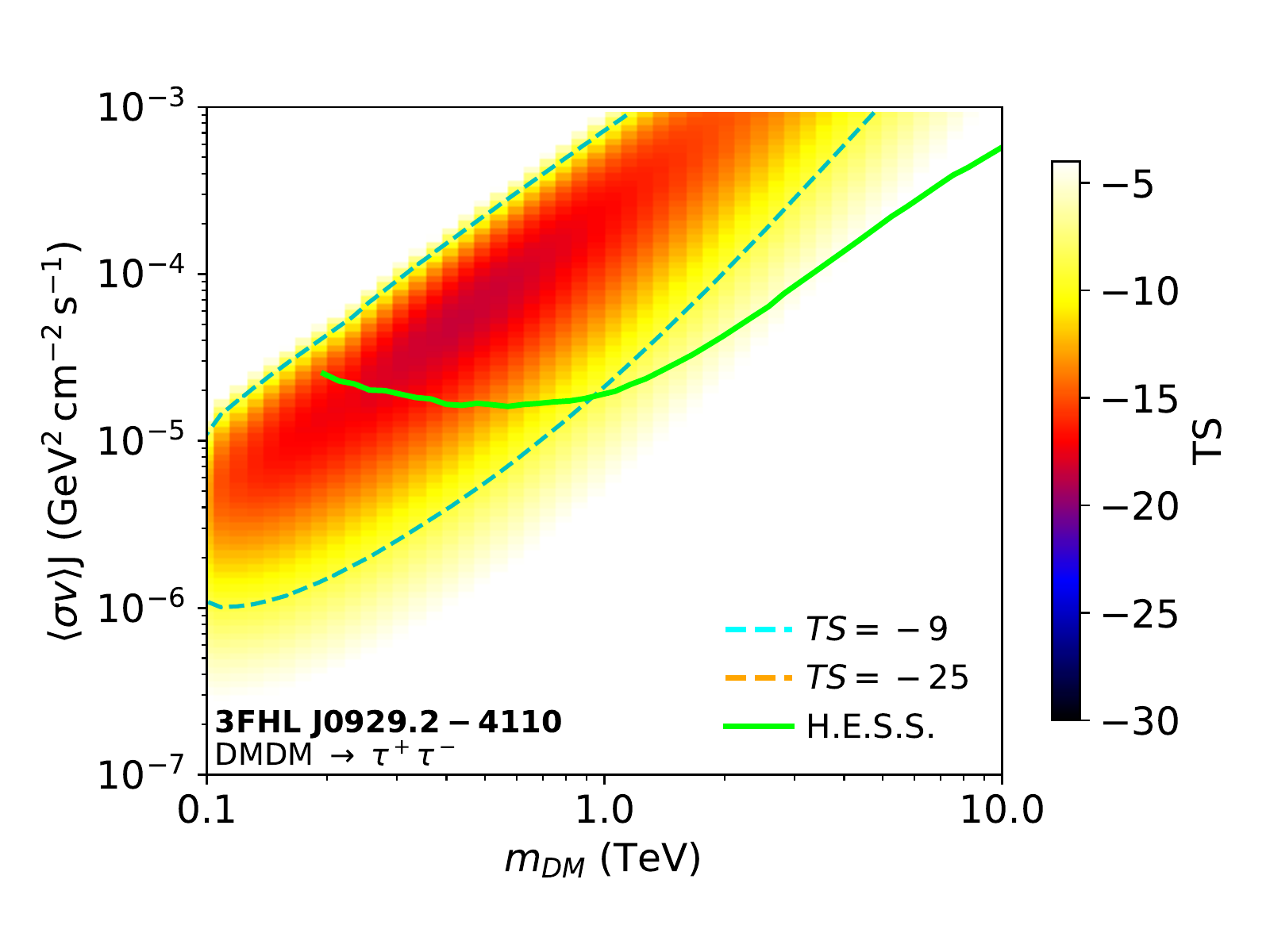}
\includegraphics[width=0.48\textwidth]{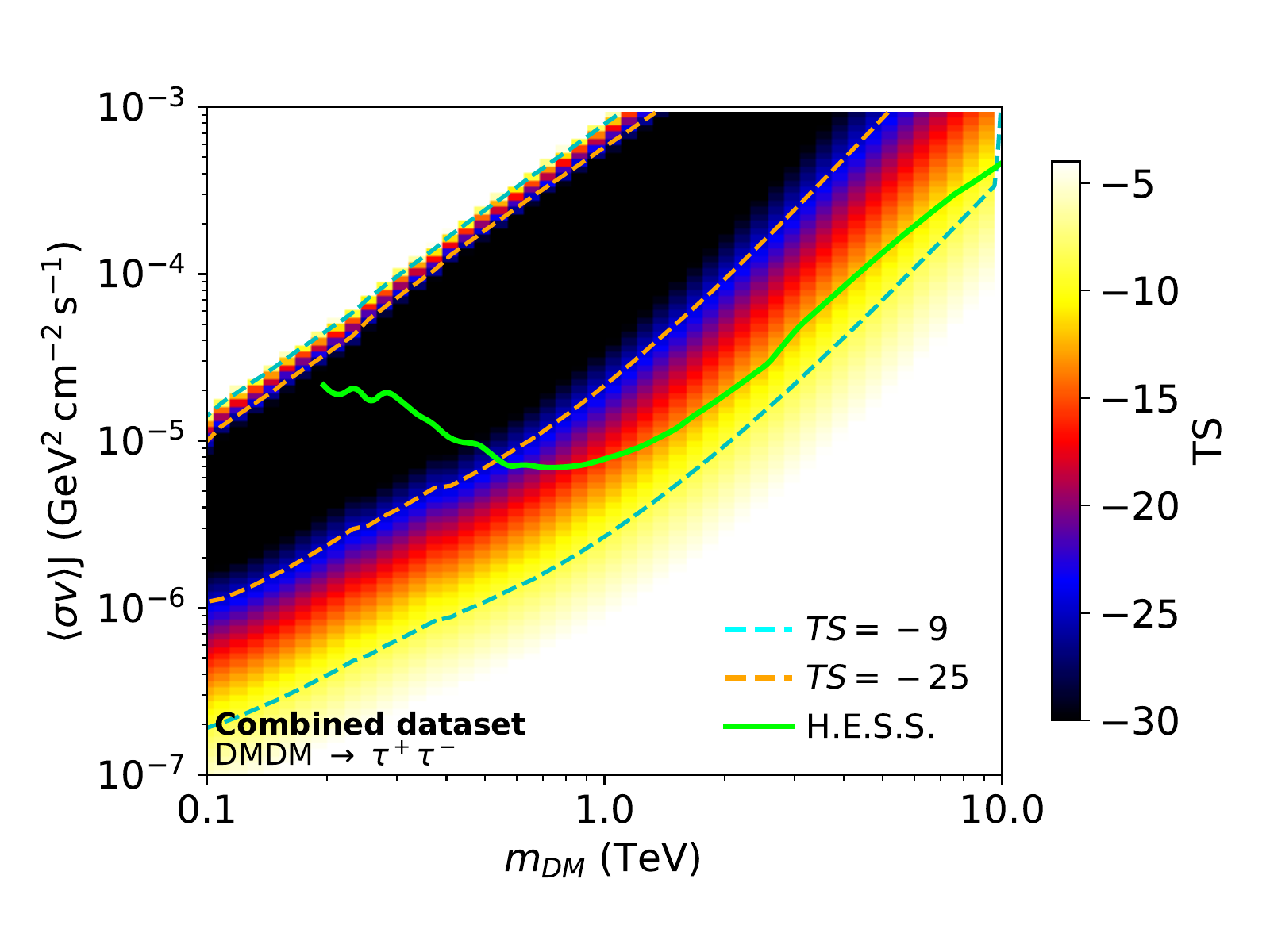}
\caption{Contours of $TS$ computed from \flat datasets on the 3FHL J0929.2-4110 (left panels) and the combined UFO datasets (right panels), respectively. The contours are given in the ($\langle \sigma v \rangle J$,m$_{\rm DM}$) plane for the $W^+W^-$ (upper panels) and $\tau^+\tau^-$ (lower panels) annihilation channel. The cyan and orange dashed lines show the $-$9 and $-$25 $TS$ contours. Overlaid (solid green line) are H.E.S.S. upper limits displayed at 95\% C.L. Contours of TS for 3FHL J1915.2-1323, 3FHL J2030.2-5037 and 3FHL J2104.5+2117  are shown in Fig.~\ref{fig:FermiTSmap1} in Appendix.}
\label{fig:FermiTSmap}
\end{figure*}

 The left panels of Fig.~\ref{fig:FermiTSmap} illustrate the results for a single UFO 3FHL~J0929.2-4110 for $W^+W^-$ (top) and $\tau^{+}\tau^-$ (bottom) annihilation channels, respectively. For each $\langle \sigma v\rangle J$, the color scale shows TS values.
 Assuming that the TS follows a $\chi^2$ distribution,  a TS equal to $-$9, (resp. $-$25) would corresponds to a 3$\sigma$  (resp. 5$\sigma$) detection for 1 degree of freedom. The dashed cyan and orange lines shows the detection region 
that corresponds to the improvement of $TS$ by $-$9 and $-$25, respectively.
The right panels in Fig.~\ref{fig:FermiTSmap} present the  results for the analysis of the combined datasets of the three selected UFOs (3FHL J0929.2-4110 ; 3FHL J1915.2-1323; 3FHL J2030.2-5037) obtained through the combination of the log-likelihood profiles from individual objects, for $W^+W^-$ (top) and $\tau^{+}\tau^-$ (bottom) annihilation channels, respectively. 

\section{H.E.S.S observations and analysis}
\label{sec:analysis}
\subsection{Data analysis}
H.E.S.S. is an array of five IACTs located in the Khomas Highland in Namibia, at an altitude of 1800 m. The array is composed of four 12~m diameter telescopes (CT1-4)  and a fifth 28~m diameter telescope (CT5) at the middle of the array. 
The observations presented here were performed in 2018 and 2019 with the full five-telescope array for the selection of unidentified Fermi-LAT objects presented in Tab.~\ref{tab:table2}. They were carried out 
in the {\it wobble} mode, where the telescope pointing direction is offset from the nominal target position by an angle between 0.5$^\circ$ and 0.7$^\circ$. The observations for the data analysis are selected  according to the standard run selection criteria~\citep{Aharonian:2006pe}.
After the calibration of raw shower images recorded in the camera, the reconstruction of the direction and energy of the gamma-ray events is performed in a template-fitting technique~\citep{2009APh32231D} in which the recorded images are compared to pre-calculated showers computed from a semi-analytical model. An energy resolution of 10\% and an angular resolution of 0.06$^\circ$ at 68\% containment radius for gamma-ray energies above 200~GeV are achieved. The results described here have been cross-checked with an independent calibration and analysis chain yielding compatible results~\citep{Parsons:2014voa}.

The selected UFOs are assumed to be pointlike sources according to the point spread function (PSF) of \flat which reaches $\sim$0.1$^\circ$ above 100 GeV. Given the H.E.S.S. PSF, the region of interest (ROI), hereafter referred to as the ON source region, is therefore defined as for pointlike-emission searches for H.E.S.S. and the ROI is taken as a disk of $0.12^\circ$ radius. The residual background is measured in OFF regions according to the {\it MultipleOff} technique~\citep{Aharonian:2006pe}. For each telescope pointing position, the OFF regions are defined at the same distance from the pointing position as for the ON region, which leads to identical acceptances in the ON and OFF regions. An excluded region defined as a disk of radius equal to twice the ON-region radius is used in order to avoid any leakage of the searched signal into the OFF regions. 
The $\alpha$ parameter is defined as the ratio between the solid angle size of the OFF and ON regions by $\alpha=\Delta\Omega_{\rm OFF}/\Delta\Omega_{\rm ON}$. The excess significance in the ROI is computed following the statistical approach of Ref.~\citep{1983ApJ...272..317L}. Table~\ref{tab:table3} summarizes for each UFO the live time, the mean zenith angle of the observations, the ON and OFF counts, the $\alpha$ parameter averaged over all the observations, as well as the excess significance in the ROI. No significant gamma-ray excess is found, neither in the ON source region, nor anywhere else in the field of view. 
\begin{table}[!htbp]
\centering
{\scriptsize
\begin{tabular}{ l | c | c | c | c | c | c }
\hline
\hline
Name &  Live time  & Mean zenith angle  & N$_{\rm ON}$ & N$_{\rm OFF}$ & $\bar{\alpha}$ & Significance\\
            &[hours] & [degrees] &[counts] & [counts] & & $[\sigma]$\\
\hline
3FHL J0929.2-4110 & 27.4 & 29.0 & 424 & 5884 & 13.9 & 0.1 \\
3FHL J1915.2-1323 & 3.6 & 19.4 & 87 & 1181 & 13.9 & 0.2 \\
3FHL J2030.2-5037 & 9.8 & 31.3 & 160 & 2192 & 13.9 & 0.1 \\
3FHL J2104.5+2117 & 6.8 & 46.7 & 73 & 853 & 13.9 & 1.1 \\
\hline
\hline
\end{tabular}
}
\caption{\label{tab:table3} H.E.S.S. data analysis results for each UFO. The second and third columns give the live time and mean zenith angle of the H.E.S.S. observations, respectively. Count numbers measured in the ON and OFF regions are given in the fourth and fifth columns, respectively, with the $\alpha$ parameter averaged over all observations, $\bar{\alpha}$, given in the sixth column. The seventh column provides the measured excess significance between the ON and OFF counts.}
\end{table}

Since no significant excess is found from any of the selected UFOs, energy-differential observed flux upper limits are computed at 95\% C.L. assuming the best-fit power-law spectral index derived from the \flat data analysis.
Expectations for the energy-differential flux upper limits are computed from 100 independent Poisson realizations of the measured OFF counts for the ON and OFF regions. From the realizations, mean and standard deviation values are extracted and used to compute the 68\% and 95\% containment bands. The differential flux upper limits are shown in Fig.~\ref{fig:sed} with an energy binning of 0.5 dex. The expectations and the containment bands for the upper limits shown in the figure are computed including 25\% systematic uncertainty. Assuming a power-law spectral index of 2 would change the differential flux upper limits by 
less than 6\%.

\subsection{Upper limit computation for the DM emission}
A binned Poisson maximum likelihood analysis is performed to 
search for spectral features expected from DM annihilation signals with respect to the background. 
For each UFO, the H.E.S.S. energy range is divided into 
62 logarithmically-spaced bins from 100 GeV up to 70 TeV.
For a given DM mass and annihilation channel, the Poisson likelihood function in the energy bin $i$ can be written as
\begin{equation}
\mathcal{L}_{\rm i}(N^{\rm S},N^{\rm B}|N_{\rm ON},N_{\rm OFF},\alpha) = \frac{(N_{\rm i}^{\rm S}+N_{\rm i}^{\rm B})^{N_{{\rm ON, i}}}}{N_{{\rm ON, i}}!}e^{-(N_{\rm i}^{\rm S}+ N_{\rm i}^{\rm B})} 
\frac{(N_{\rm i}^{\rm S'}+\alpha N_{\rm i}^{\rm B})^{N_{{\rm OFF, i}}}}{N_{{\rm OFF, i}}!}e^{-(N_{\rm i}^{\rm S'}+\alpha N_{\rm i}^{\rm B})} \, .
\label{eq:likelihood}
\end{equation}
$N_{\rm ON, i}$ and $N_{\rm OFF, i}$ stand for the number of measured events in the ON and OFF regions, respectively. 
$N^{\rm B}_{\rm i}$ is the expected number of background events in the ON region. 
$N^{\rm S}_{\rm i}$ and $N^{\rm S'}_{\rm i}$ are the expected number of DM signal events in the ON and OFF regions, respectively.
They are computed by folding the expected theoretical DM flux given in Eq.~(\ref{eq:dmflux}) with the energy-dependent acceptance and energy resolution of H.E.S.S. for the considered data set. The term $dN/dE_{\gamma}^{f}$ in Eq.~(\ref{eq:dmflux}) is extracted from Ref.~\citep{Cirelli:2010xx} for each assumed DM mass and annihilation channel.
The energy resolution of H.E.S.S. is represented by a Gaussian function with a width of $\sigma_{\rm E}/E$ = 10\% above 200 GeV.  
 The UFOs are also pointlike sources for H.E.S.S., therefore no leakage is expected in the background region and $N^{\rm S'}_{\rm i}$ is taken to $N^{\rm S'}_{\rm i} \equiv$ 0. The likelihood function for a given object $\mathcal{L}$ is defined as $\mathcal{L} = \prod_{\rm i}\mathcal{L}_{\rm i}$.
 
Since no significant excess is found in any of the selected UFOs by H.E.S.S., upper limits can be derived assuming UFOs are DM-induced gamma-ray emitters from a likelihood ratio test statistic (TS) given by: 
\begin{equation}
TS = - 2\, {\rm \log} \frac{\mathcal{L}(N^{\rm S}(\langle \sigma v\rangle J), \widehat{\widehat {N^{\rm B}}}(\langle \sigma v\rangle J)|N_{\rm ON},N_{\rm OFF},\alpha)}{\mathcal{L}(\widehat{N^{\rm S}}(\langle \sigma v\rangle J),\widehat{N^{\rm B}}|N_{\rm ON},N_{\rm OFF},\alpha)} \, .
\label{eq:TS}
\end{equation}
$\widehat{\widehat {N^{\rm B}_{\rm i}}}$ 
is obtained through a conditional maximization, achieved by solving
$d\mathcal{L}/dN^{\rm B}_{\rm i} = 0$. $\widehat{N^{\rm S}_{\rm i}}$ and $\widehat{N^{\rm B}_{\rm i}}$ are computed using an unconditional maximization. Following the procedure defined in Ref.~\citep{2011EPJC711554C}, upper limits are computed assuming a positive signal, {\it i.e.}, $\widehat{N^{\rm S}_{\rm i}}>0$.
The $\langle \sigma v \rangle J$  value for which the TS value is equal to 2.71 provides the one-sided 95\% confidence level (C.L.) upper limit on the quantity $\langle \sigma v \rangle J$.

The hypothesis that all UFOs are indeed DM subhalos, but too faint to be detected as such in the TeV energy range with the given exposure, can be tested when the individual datasets are combined.  
If no significant overall excess is found in the combined dataset, combined upper limits on $\langle \sigma v \rangle J$ can be derived versus the DM mass  assuming $J$ to be an averaged $J$-factor. In Eq.~(\ref{eq:TS}), the likelihood function is replaced by the combined likelihood expressed as $\mathcal{L}_{\rm comb} = \prod_{j=1}^{\rm N_{targets}} \mathcal{L}_{\rm j}$, where $\mathcal{L}_{\rm j}$ is the likelihood of the target $j$.

\section{Results}
\label{sec:results}
\begin{figure*}[htbp]
\centering
\includegraphics[width=0.45\textwidth]{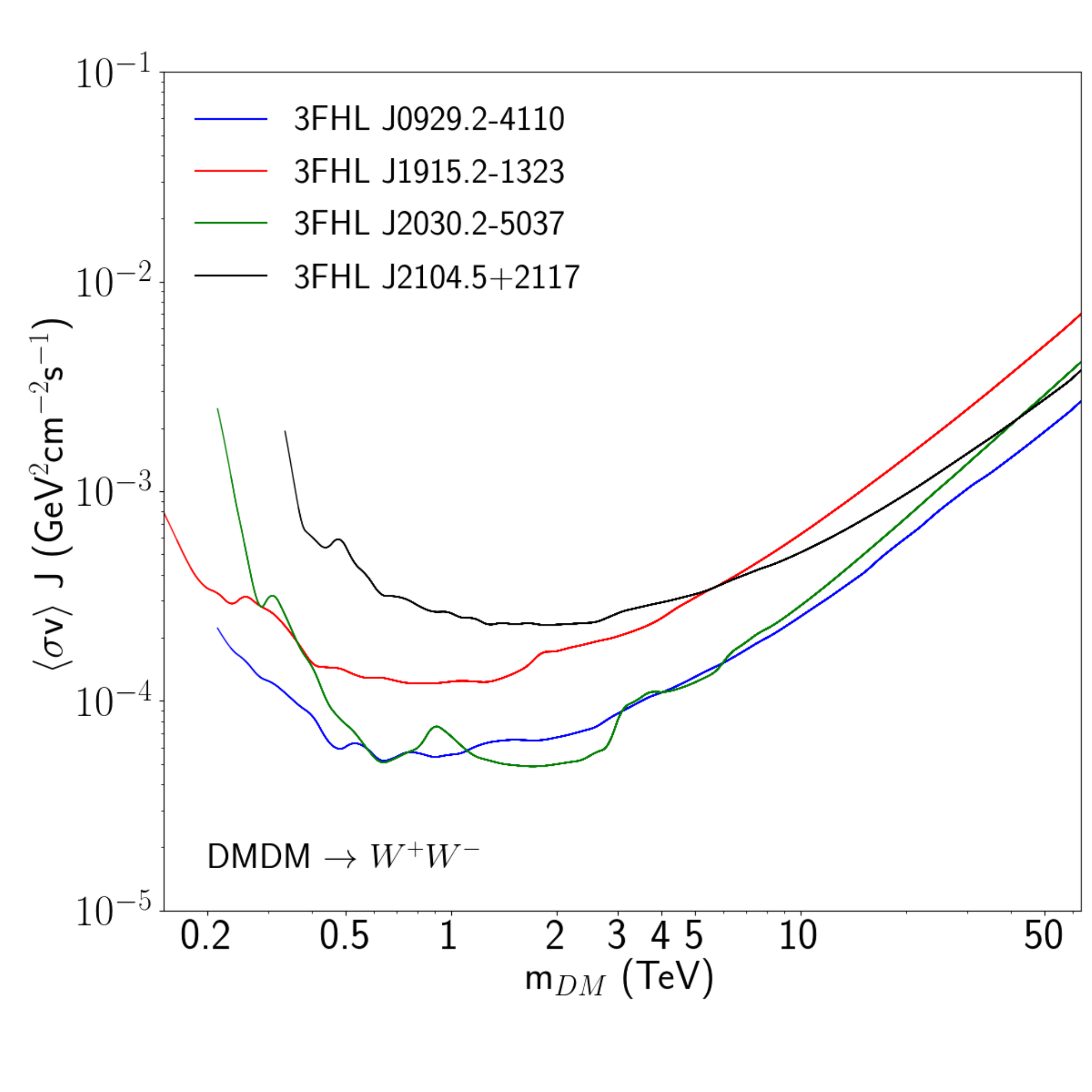}
\includegraphics[width=0.45\textwidth]{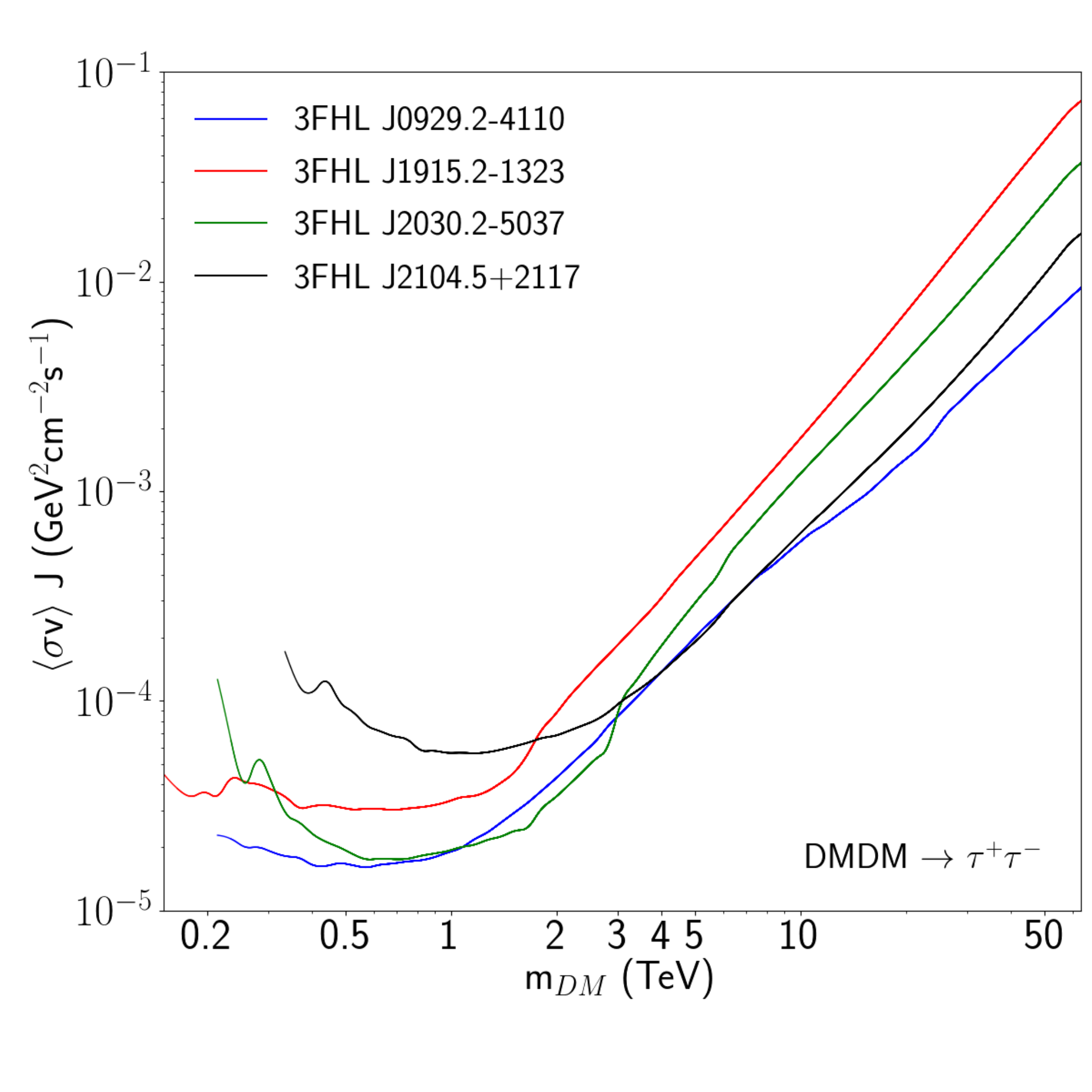}
\caption{95\% C.L. upper limits on the product of the annihilation cross section $\langle\sigma v\rangle$ and the $J$-factor $J$ as a function of the DM mass $m_{\rm DM}$ in the $W^+W^-$ (left panel) and $\tau^+\tau^-$ (right panel) annihilation channels for 3FHL J0929.2-4110 (blue line), 3FHL J1915.2-1323 (red line), 3FHL J2030.2-5037 (green line), and 3FHL J2104.5+2117 (black line), respectively.}
\label{fig:sigmavJ_all}
\end{figure*}

No significant excess is measured in any of the H.E.S.S. datasets of the selected UFOs. 95\% C.L. upper limits 
on $\langle \sigma v \rangle J$ are derived versus the DM mass 
using Eq.~(\ref{eq:TS}).
Figure~\ref{fig:sigmavJ_all} shows for each UFO the upper limits as a function of the DM mass for the $W^+W^-$ 
and $\tau^+\tau^-$ annihilation channels, respectively.
In most of the DM mass range , the strongest constraints are reached for 3FHL J0929.2-4110 observations. For a 1 TeV DM mass, the constraints $\langle \sigma v \rangle J$ = 5.5$\times$10$^{-5}$ GeV$^2$cm$^{-2}$s$^{-1}$ and 1.9$\times$10$^{-5}$ GeV$^2$cm$^{-2}$s$^{-1}$ in $W^+W^-$ and $\tau^+\tau^-$ annihilation channels, respectively, for 3FHL J0929.2-4110.

The combined analysis of the four H.E.S.S. datasets does not show any significant excess. Therefore, the UFO datasets are combined and upper limits on $\langle \sigma v\rangle J$ are computed. Given its possible association with an AGN, the source 3FHL J2104.5+2117 is removed to provide conservative combined upper limits. 
 The right panel of Fig.~\ref{fig:FermiTSmap} shows the combined 95\% C.L. upper limits on $\langle \sigma v\rangle J$ as a function of the DM mass for the $W^+W^-$ and  $\tau^+\tau^-$
 annihilation channels, respectively. The analysis of the combined datasets allows for an improvement of about 10\% and 20\% for 1 TeV DM mass in the $W^+W^-$ and $\tau^+\tau^-$ annihilation channel, respectively,
 with respect to the most constraining upper limit from the
 individual UFO datasets. 
The combined limits reach 3.7$\times$10$^{-5}$ and 8.1$\times$10$^{-6}$ GeV$^2$cm$^{-2}$s$^{-1}$ in the $W^+W^-$ and $\tau^+\tau^-$ channels, respectively, for a 1~TeV DM mass.
\begin{figure*}[htbp]
\centering
\includegraphics[width=0.48\textwidth]{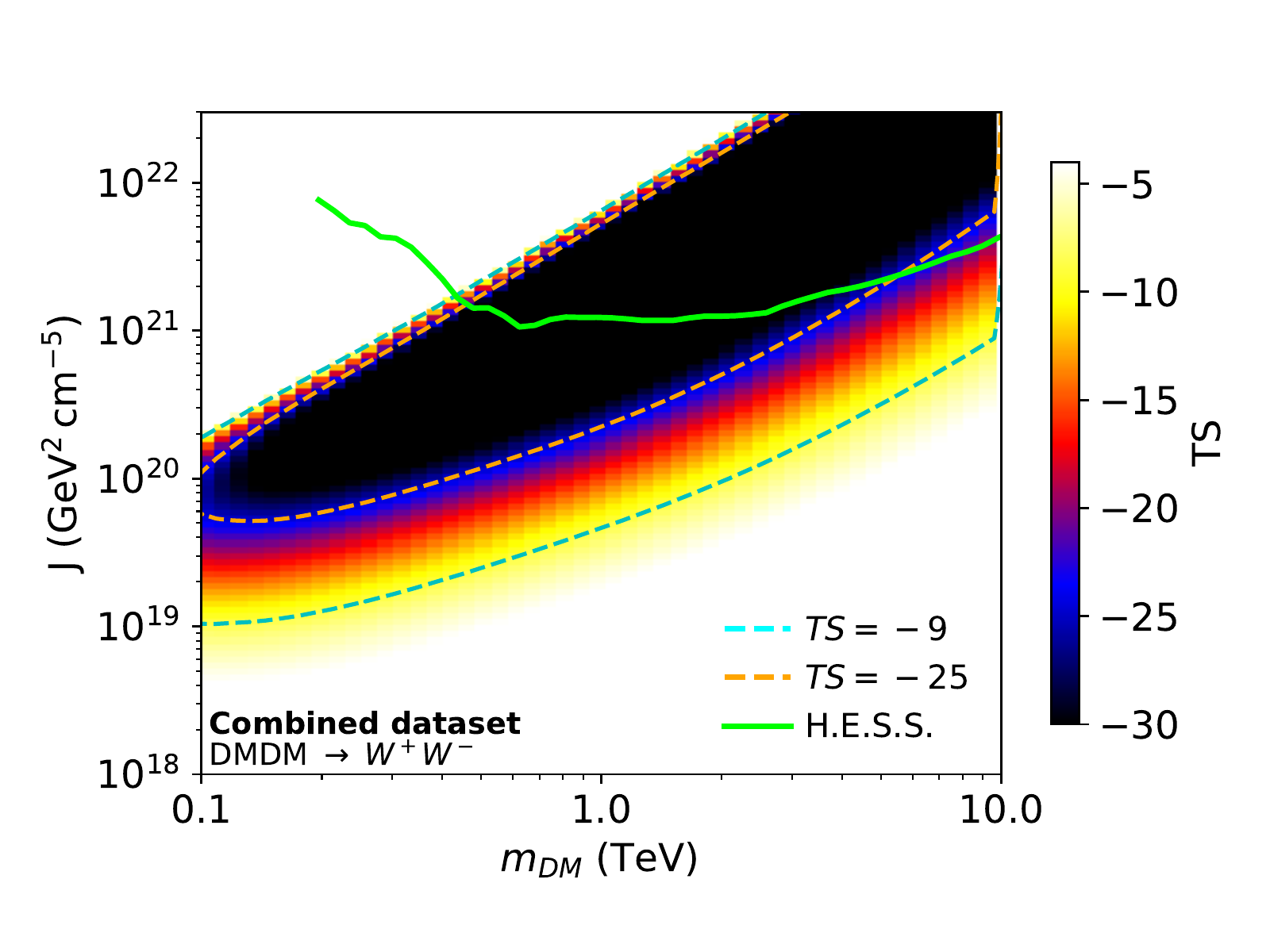}
\includegraphics[width=0.48\textwidth]{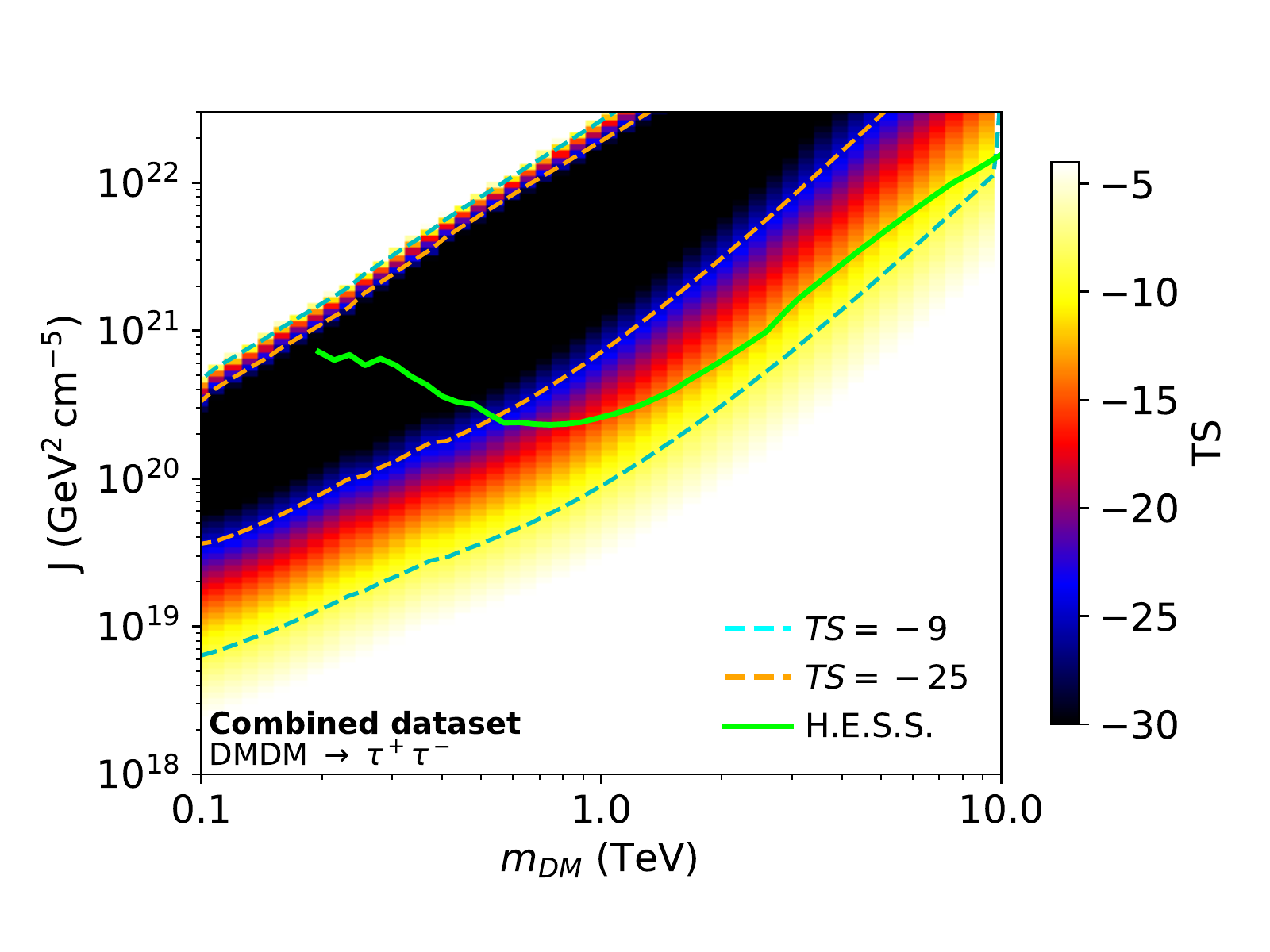}
\caption{ Contours of $TS$ computed from the \flat combined UFO datasets. The contours are given in the ($J$,m$_{\rm DM}$) plane for the $W^+W^-$ (left panel) and $\tau^+\tau^-$ (right panel) annihilation channel,
assuming the $\langle\sigma v\rangle$ value expected for thermal WIMPs.
The cyan and orange dashed lines show the $-$9 and $-$25 $TS$ contours. Overlaid (solid green line) are the H.E.S.S. 95\% C.L. upper limits from the combined UFO datasets. }
\label{fig:sigmavJ_stacked}
\end{figure*}

In order to derive the $J$-factor values required to explain the UFO emission in terms of DM models, the value of the annihilation cross section expected for thermal WIMPs ($\langle\sigma v\rangle_{\rm th} \simeq$ 3$\times$10$^{-26}$ cm$^3$s$^{-1}$) is used~\citep{Steigman:2012nb}.
The $J$-factor upper limits for the DM models of the UFOs as function of the DM mass are given at 95\% C.L. in  Fig.~\ref{fig:sigmavJ_stacked}.
For a 1~TeV DM mass in the $W^+W^-$ channel, the $J$-factor values are constrained to be between 2.4$\times$10$^{20}$ and 1.3$\times$10$^{21}$ GeV$^2$cm$^{-5}$ for DM models with $TS \le$ $-$25 (which corresponds to $\ge$5$\sigma$ confidence interval assuming $TS$ follows $\chi^2$ distribution). For a DM mass of 10 TeV in the $W^+W^-$ channel, all the $J$-factor values for DM models with TS $\le$ $-$25 are ruled out at 95\% C.L. by the H.E.S.S. constraints. In the $\tau^+\tau^-$ channels, the H.E.S.S. constraints are even stronger. For 
300~GeV DM mass, the allowed $J$-factor values are between 1.4$\times$10$^{20}$ and 5.9$\times$10$^{20}$ GeV$^2$cm$^{-5}$ for TS $\le$ $-$25 DM models. The H.E.S.S. upper limits restrict the J-factors to lie in the range $6.1\times 10^{19} -2.0\times 10^{21}$~GeV$^2$cm$^{-5}$ and the masses 
to lie between 0.2 and 6 TeV in the W$^+$W$^-$channel. For the $\tau^+\tau^-$ channel, the J-factors lie in the range $7.0\times 10^{19} - 7.1\times 10^{20}$~GeV$^2$cm$^{-5}$ and the masses lie between 0.2 and 0.5~TeV.

Using predictions of N-body cosmological simulations, the number of subhalos with a $J$-factor higher than a given value for a MW-like galaxy can be extracted as displayed in Fig.~\ref{fig:luminosty_function}. The probability to have at least three subhalos with a $J$-factor higher than 10$^{20}$ GeV$^2$cm$^{-5}$ is below 5\% (Fig.~\ref{fig:luminosty_function}, blue-dotted line). 
According to this prediction, the interpretation of the UFO emissions in terms of DM particle annihilations in Galactic DM subhalos can be further
constrained from Fig.~\ref{fig:sigmavJ_stacked} to $m_{\rm DM}\lesssim 1$~TeV for $W^+W^-$ and $m_{\rm DM}\lesssim 0.3$~TeV for $\tau^+\tau^-$ channels.

\section{Discussion}
\label{sec:discussion}
A substantial number of UFOs may emit gamma rays from DM annihilations in subhalos. However, some of them could be active galactic nuclei or other types of galaxies that lack so far detection at other wavelengths. Alternative astrophysical interpretations of UFOs as  pulsars or low-luminosity globular clusters hosting millisecond pulsars~\citep{Mirabal:2016huj}, may be less plausible since typical gamma-ray spectra for these types of objects are characterised by an energy cut-off at energies of a few GeV.

The cumulative J-factor distribution is in very good agreement with the results of~\cite{hutten16} for the "HIGH" model intended to predict the highest possible number of subhalos in a typical MW-like galaxy. 
The real number of DM subhalos can be an order of magnitude smaller as shown for the predictions in the ``LOW'' model of~\cite{hutten16}.
The choice of the number of subhalos of masses between 10$^8$ and 10$^{10}$ M$_\odot$ of $N_{\rm calib}$ = 300 is motivated by the output of DM-only simulations~\cite{Springel:2008cc}.
Baryon feedback can significantly reduce this value, up to a factor of two~\citep{Mollitor:2014ara,Sawala:2015cdf}. This would make 
the highest J-factor values even more unlikely.
As discussed in~\cite{Coronado-Blazquez:2019pny},  subhalos with the highest $J$-factors should appear as extended sources for \flat given its point spread function of about 0.1$^\circ$ above 10~GeV. 
However, even these brightest DM subhalos would produce faint gamma-ray sources whose spatial extension would be challenging to measure for \flat. On the simulation front, 
further work is likely needed to use predictions for subhalo angular sizes in MW-like galaxies to definitely rule out pointlike
UFOs as potential DM subhalos.

The DM density distribution of the Galactic halo is assumed here to follow a NFW parametrization.
Incorporating hydrodynamics and baryon feedback in cosmological simulations tend to soften the inner cusp of the DM profiles in Milky Way-like galaxies, 
leading to a flattening of order 1 kpc~\citep[see, for instance,][]{Chan:2015tna}.
However, these predictions for the expected DM distribution have large uncertainties due to the effects of baryonic physics. The resolution limit of the simulations at sufficiently small distances becomes also pertinent.
Alternative Galactic mass models can be used to describe subhalo parameters for Milky Way-like galaxies~\citep{Catena:2009mf,2011MNRAS.414.2446M,Stref:2016uzb,2017MNRAS.465...76M}. Using different Galactic mass models, the subhalo luminosity functions derived  in~\cite{Stref:2016uzb} provide compatible results.
Considering a core profile  
would make the high DM mass exclusion of the DM models for the UFO emission even stronger. DM subhalos could also have cored profiles with lower DM concentration, which would make them more prone to tidal disruption. This would therefore decrease both the normalisation of the $J$-factor distribution function and shift it to lower values. 
The $J$-factor distribution prediction is based on DM-only Via Lactea-II simulations using WMAP cosmology. 
The simulations based on the 
most recent cosmological results from the Planck mission
with added baryonic physics could somewhat change the predicted properties of the MW subhalos.
The effect of baryon feedback and tidal effects induced by both DM and baryons is likely to alter the DM concentration of subhalos~\citep{Despali:2016meh,Stref:2016uzb}.
Details of the modeling
of tidal disruption of Galactic DM subhalos on the brightest subhalo can be found in \cite{DiMauro:2020uos}.
Therefore, the cumulative $J$-factor distribution would be shifted to lower values. In this case, the DM-induced interpretation of the UFO emission would be even more constrained given that the probability to find high $J$-factor values would be even smaller.
The presence of baryons affects both the amplitude and the slope of the DM halo and subhalo mass functions. 
It can reduce the slope by a few percent in the 10$^6$ - 10$^9$ M$_\odot$ subhalo mass range~\citep{Benson:2019jio}, which can also 
alter the rate of subhalos with large J factors.   No cut is applied to the maximal value of the subhalo mass when computing the cumulative J-factor distribution. While subhalos with masses of about 10$^7$ M$_\odot$ or higher may be able to host star formation and actually be dwarf galaxies, simulations including
hydrodynamics and feedback physics in addition to the 
gravitational effects for the expected DM distribution in both the main halo and its subhalos
such as in~\cite{Zhu:2015jwa}, show that a significant fraction of subhalos with masses of about 10$^9$ M$_\odot$ is found to host no stars. For subhalo masses larger than 10$^7$-10$^8$ M$_\odot$, a noticeable fraction of them can start triggering star formation and indeed form a faint dwarf galaxy. Arguably these masses  critically depend on the baryonic physics implementation in the simulations and its associated feedback.
Considering subhalos with masses lower than 10$^7$ M$_\odot$ implies the probability to have at least one subhalo with J $\ge$ 3$\times 10^{20}$~GeV$^2$cm$^{-5}$ to be about 0.3\%, therefore a factor of about 16 lower that in the case without mass cut.

The interpretation of UFOs as DM subhalos of TeV-mass scale thermal WIMPs requires $J$-factors in excess of a few 10$^{20}$~GeV$^2$cm$^{-5}$.
Such $J$-factor values are only occasionally obtained in N-body simulations of MW-type galaxies. The highest subhalo $J$-factor values are usually subject to a large statistical variance. The precise value of the brightest subhalos can be subject to a large uncertainty, implying a factor-of-ten uncertainty on the $J$-factor value for J $\gtrsim$ 10$^{20}$ GeV$^2$cm$^{-5}$  in the "HIGH" model~\citep{hutten19}\footnote{
Using the "LOW" model for the predictions of the J-factor distribution would make the interpretation of UFOs as DM subhalos even more unlikely given that the probability to get high J-factor values would be lowered compared to what it would be in the "HIGH" model.}. 
Previous studies using \flat data only searched for UFOs as promising DM subhalo candidates for DM masses below 100 GeV~\citep[see, for instance,][]{Bertoni:2015mla,Bertoni:2016hoh,Coronado-Blazquez:2019puc}. For instance, in the study of \cite{Coronado-Blazquez:2019pny}, masses of few ten GeV are ruled out for canonical thermal WIMPs.
The present analysis searches for UFOs as DM subhalos for the uncharted TeV-mass thermal WIMP models.

The maximal value of subhalo $J$-factors obtained in simulations is model dependent and can be increased even in comparison to the optimistic estimate of the $J$-factor distribution considered here as discussed below.
The normalisation of the subhalo mass function assumes usually about 10\% of the total DM halo content to be in the form of subhalos, with the total DM halo density being normalized to the DM density at the location of the Sun $\rho(r_\odot)$ = $\rho_\odot$ = 0.39 GeVcm$^{-3}$. The precise value is subject to uncertainties of a factor of 1.5 to 2~\citep{Read:2014qva}. 
 Varying the input parameters of the simulations in the relevant ranges of interest, such as $\rho_\odot$ = 0.6 GeVcm$^{-3}$ and the scale radius of the main DM halo $r_{\rm s}$ = 25 kpc, would increase the highest $J$-factor values by a factor of two. The predicted cumulative $J$-factor distribution
would therefore probe higher $J$-factor values.
Considering substructures in galactic subhalos (see, for instance, \cite{Hiroshima:2018kfv}) results in higher expected $J$-factor values for the Galactic subhalo population 
with typical increase factors of a few, therefore shifting the cumulative J-factor distribution to higher values.

The above-mentioned large systematic uncertainties in the prediction of the $J$-factor distribution 
 weaken significantly the constraints from cosmological simulations, which makes them comparable to or weaker than the H.E.S.S. constraints in, {\it e.g.}, the $\tau^+\tau^-$ channel. 
This makes the model-independent H.E.S.S. constraints the only relevant ones for robust interpretation of UFOs as Galactic subhalos of annihilating DM.

\section{Summary}
\label{sec:summary}
In this work, a straightforward filtering of the unidentified sources in the 3FHL point-source catalogue has been performed using selection cuts to identify the most promising DM subhalo candidates for DM masses above a few hundred GeV. The datasets for the four UFOs were collected with the Fermi satellite in a 12-year observation period.
Using H.E.S.S. observations, no significant signal has been detected from any of the selected UFOs. From H.E.S.S. flux upper limits, DM models describing the UFO emissions with high significance are strongly constrained in the TeV DM mass range for different annihilation channels. Assuming thermally-produced DM particles, the DM models for the UFO emissions require high $J$-factor values. 
When the model-dependent predictions from $N$-body simulations of the MW-like subhalo population are taken into account, the required high $J$-factor values for the DM models explaining the UFOs as Galactic subhalos are 
unlikely.
This can point towards interpretation of the UFOs as subhalos of relatively light WIMPs with masses $m_{\rm DM}\lesssim 0.3$~TeV for which somewhat lower $J$-factors are preferred. 
However, this could be in tension with constraints on thermal WIMPs from dwarf galaxy observations by 
\flat ~\citep{Ackermann:2015zua,Fermi-LAT:2016uux}.

\section*{Acknowledgements}
The support of the Namibian authorities and of the University of Namibia in facilitating the construction and operation of H.E.S.S. is gratefully acknowledged, as is the support by the German Ministry for Education and Research (BMBF), the Max Planck Society, the German Research Foundation (DFG), the Helmholtz Association, the Alexander von Humboldt Foundation, the French Ministry of Higher Education, Research and Innovation, the Centre National de la Recherche Scientifique (CNRS/IN2P3 and CNRS/INSU), the Commissariat \`a l'\'energie atomique et aux \'energies alternatives (CEA), the U.K. Science and Technology Facilities Council (STFC), the Knut and Alice Wallenberg Foundation, the National Science Centre, Poland grant no. 2016/22/M/ST9/00382, the South African Department of Science and Technology and National Research Foundation, the University of Namibia, the National Commission on Research, Science \& Technology of Namibia (NCRST), the Austrian Federal Ministry of Education, Science and Research and the Austrian Science Fund (FWF), the Australian Research Council (ARC), the Japan Society for the Promotion of Science and by the University of Amsterdam.

We appreciate the excellent work of the technical support staff in Berlin, Zeuthen, Heidelberg, Palaiseau, Paris, Saclay, T\"ubingen and in Namibia in the construction and operation of the equipment. This work benefitted from services provided by the H.E.S.S. Virtual Organisation, supported by the national resource providers of the EGI Federation.

\def\aj{AJ}%
\def\actaa{Acta Astron.}%
\def\araa{ARA\&A}%
\def\apj{ApJ}%
\def\apjl{ApJ}%
\def\apjs{ApJS}%
\def\ao{Appl.~Opt.}%
\def\apss{Ap\&SS}%
\def\aap{A\&A}%
\def\aapr{A\&A~Rev.}%
\def\aaps{A\&AS}%
\def\azh{AZh}%
\def\baas{BAAS}%
\def\bac{Bull. astr. Inst. Czechosl.}%
\def\caa{Chinese Astron. Astrophys.}%
\def\cjaa{Chinese J. Astron. Astrophys.}%
\def\icarus{Icarus}%
\def\jcap{J. Cosmology Astropart. Phys.}%
\def\jrasc{JRASC}%
\def\mnras{MNRAS}%
\def\memras{MmRAS}%
\def\na{New A}%
\def\nar{New A Rev.}%
\def\pasa{PASA}%
\def\pra{Phys.~Rev.~A}%
\def\prb{Phys.~Rev.~B}%
\def\prc{Phys.~Rev.~C}%
\def\prd{Phys.~Rev.~D}%
\def\pre{Phys.~Rev.~E}%
\def\prl{Phys.~Rev.~Lett.}%
\def\pasp{PASP}%
\def\pasj{PASJ}%
\def\qjras{QJRAS}%
\def\rmxaa{Rev. Mexicana Astron. Astrofis.}%
\def\skytel{S\&T}%
\def\solphys{Sol.~Phys.}%
\def\sovast{Soviet~Ast.}%
\def\ssr{Space~Sci.~Rev.}%
\def\zap{ZAp}%
\def\nat{Nature}%
\def\iaucirc{IAU~Circ.}%
\def\aplett{Astrophys.~Lett.}%
\def\apspr{Astrophys.~Space~Phys.~Res.}%
\def\bain{Bull.~Astron.~Inst.~Netherlands}%
\def\fcp{Fund.~Cosmic~Phys.}%
\def\gca{Geochim.~Cosmochim.~Acta}%
\def\grl{Geophys.~Res.~Lett.}%
\def\jcp{J.~Chem.~Phys.}%
\def\jgr{J.~Geophys.~Res.}%
\def\jqsrt{J.~Quant.~Spec.~Radiat.~Transf.}%
\def\memsai{Mem.~Soc.~Astron.~Italiana}%
\def\nphysa{Nucl.~Phys.~A}%
\def\physrep{Phys.~Rep.}%
\def\physscr{Phys.~Scr}%
\def\planss{Planet.~Space~Sci.}%
\def\procspie{Proc.~SPIE}%
\let\astap=\aap
\let\apjlett=\apjl
\let\apjsupp=\apjs
\let\applopt=\ao
\newpage
\bibliography{bibl} 

\appendix
\section{Test-statistic maps for the UFO sources from FERMI-LAT datasets}
\label{app:ts_maps}
\begin{figure*}
    \centering
    \includegraphics[width=0.482\textwidth]{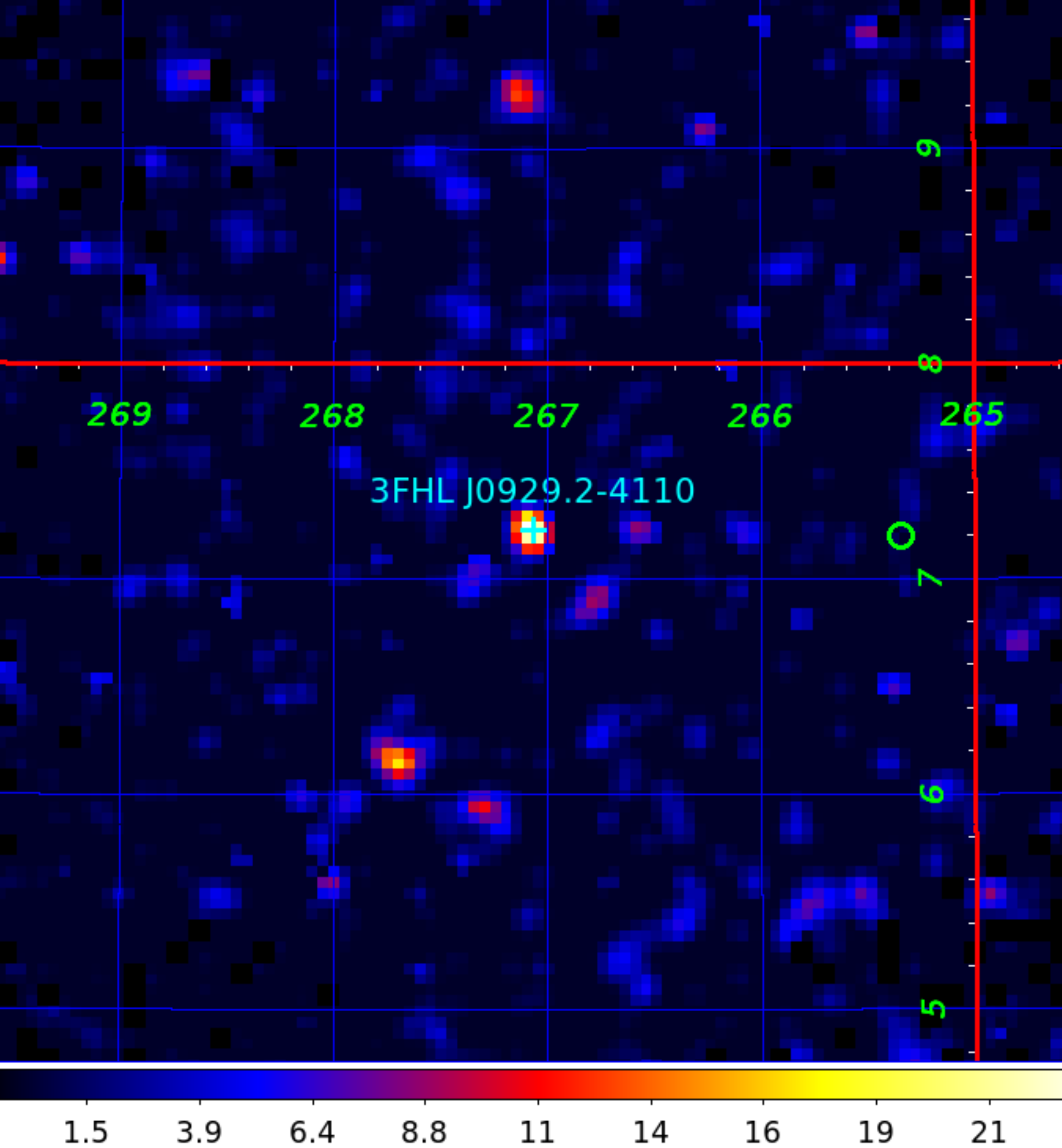}
    \includegraphics[width=0.49\textwidth]{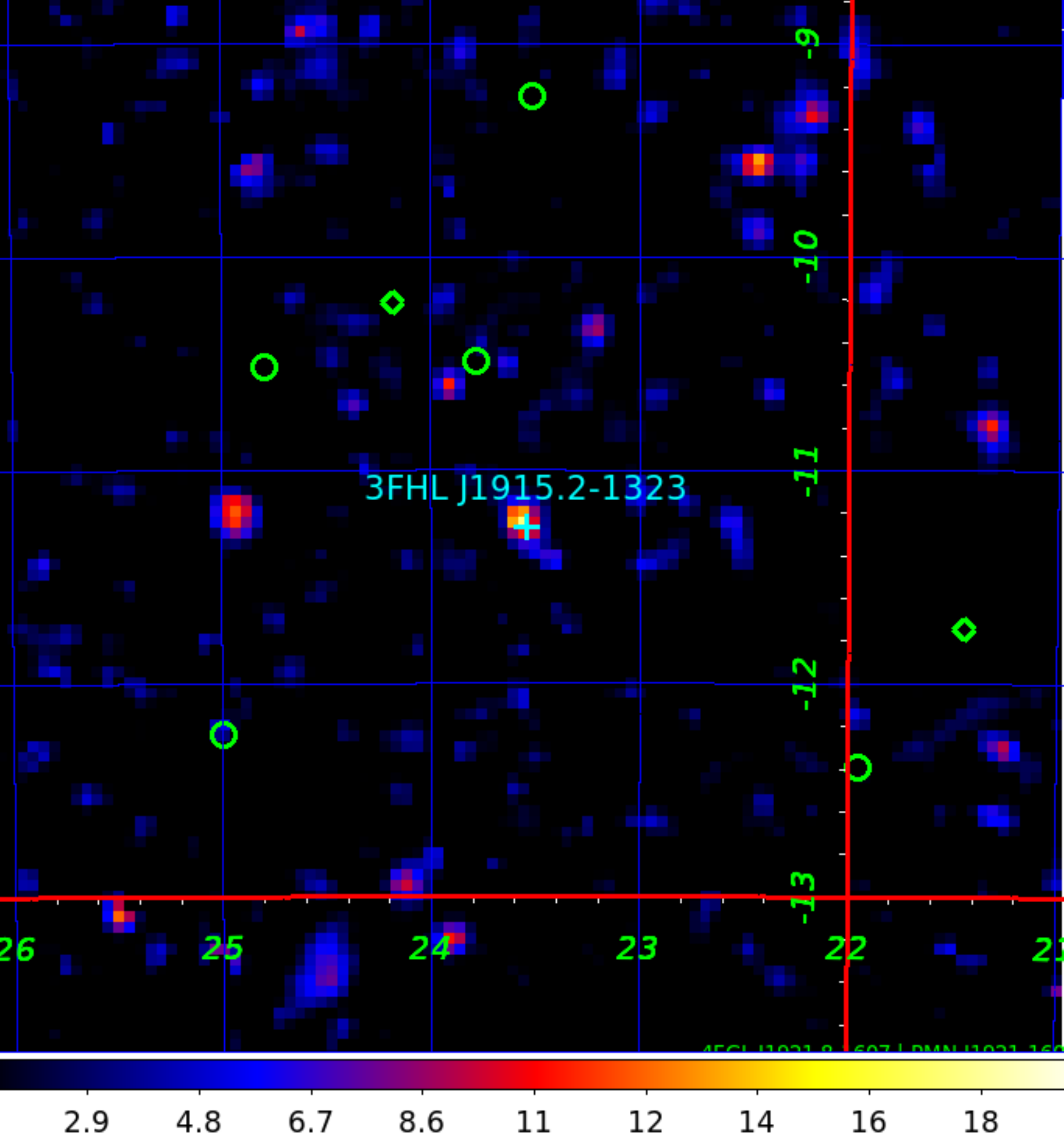}\\
    \includegraphics[width=0.49\textwidth]{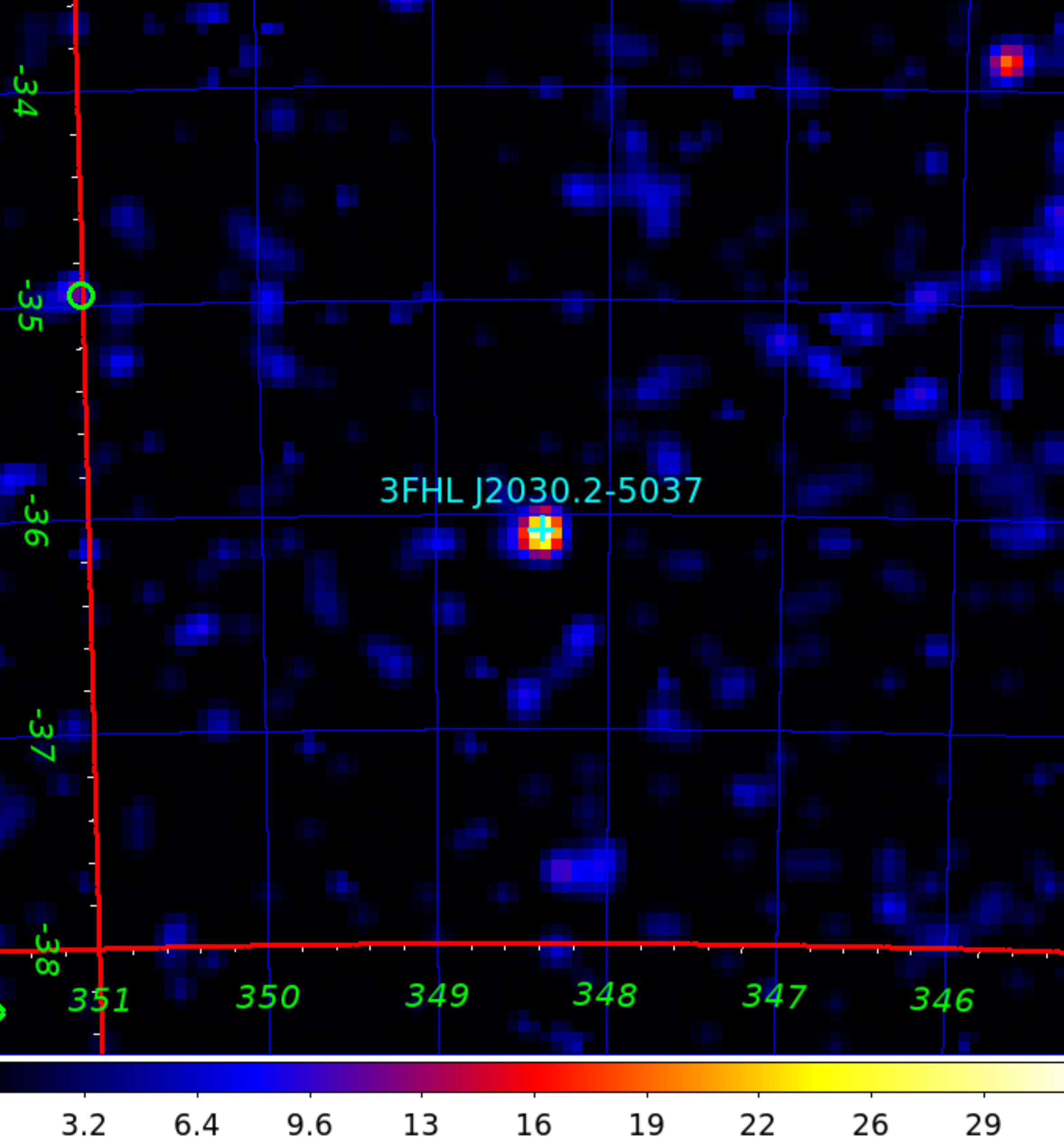}
    \includegraphics[width=0.487\textwidth]{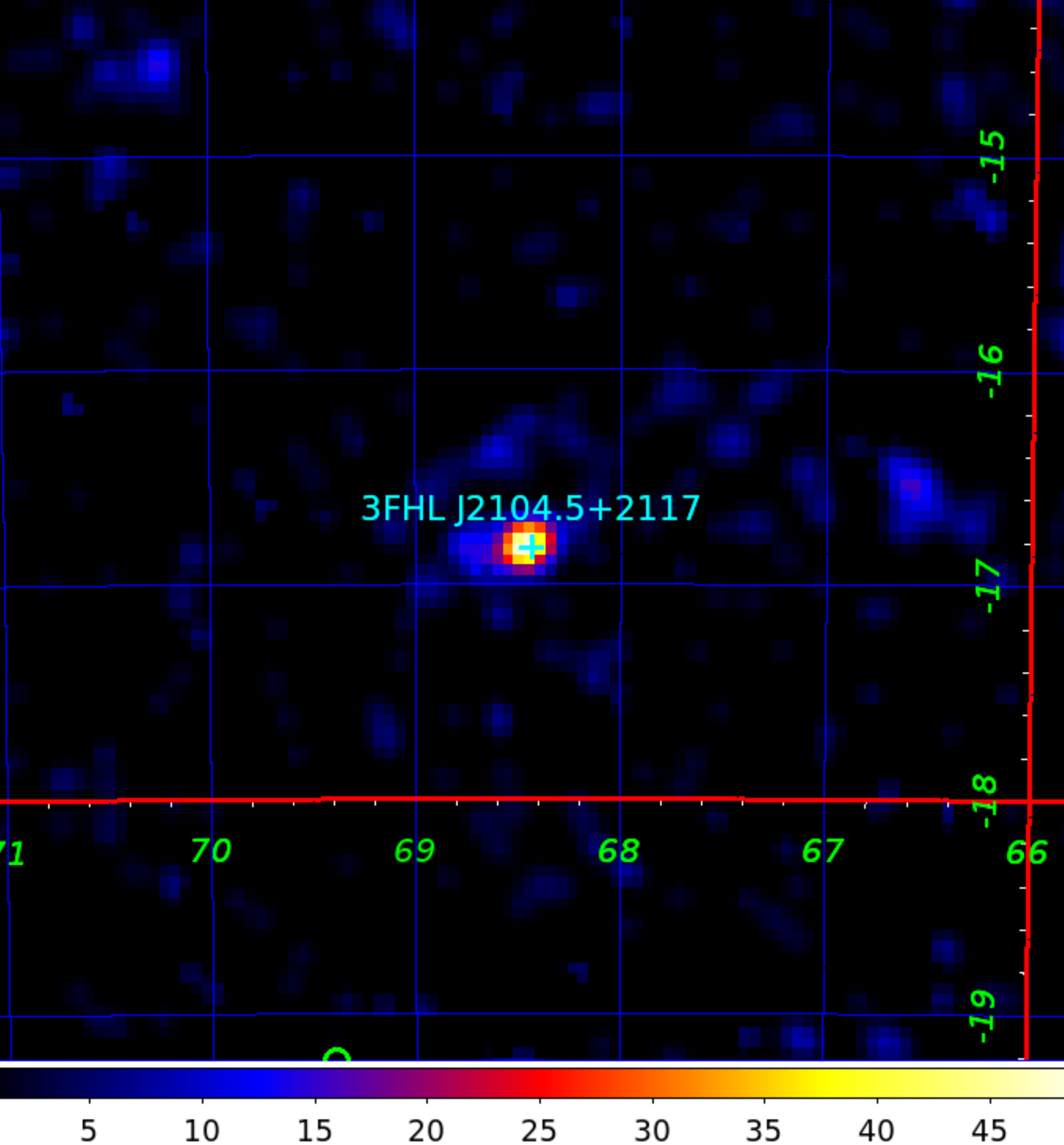}
    \caption{Test statistics maps for $5^\circ\times 5^\circ$ region around each of the considered UFOs for energies above  10~GeV in Galactic coordinates with pixel size of 0.05$^\circ$. The color scale indicates the values of the TS.
    No smoothing to the map is applied. Cyan crosses show the position of corresponding UFO source. Green crosses show positions of nearby 4FGL sources included in the background model.}
    \label{fig:ts}
\end{figure*}
In this section we present the results for the search of possible
sources not accounted by the considered model based on 4FGL catalogue of the UFOs vicinity. We built test-statistics (TS) maps for $5^\circ\times 5^\circ$ region centered at the position of each UFO. These maps illustrate the significance ($\sim \sqrt{TS}$) above the background model of a test point-like source with a power-law spectrum characterised by a free normalisation and slope fixed to -2, in each pixel.

For the background model, where the corresponding UFO source is removed, 
we consider the same spatial/spectral models as used for the analysis of \flat data described in Sec.~\ref{sec:fermi_data_analysis}. 
Such a choice of the background model allows us to check the presence of unaccounted sources close to UFOs' positions as well as verify the point-like spatial morphology of the emission associated with the UFO.

Fig.~\ref{fig:ts} show the TS maps for all four UFO sources for energies above 10 GeV in Galactic coordinates with pixel size of $0.05^\circ$. No smoothing is applied to the maps. 
 Cyan crosses show the positions of UFO sources. Green circles indicate the position of the nearby 4FGL sources included in the background model. The maps indicate the absence of significant residuals which could affect the \flat analysis of the UFOs.

\newpage
\section{Contours of TS from Fermi-LAT datasets on the 3FHL J1915.2-1323, 3FHL J2030.2-5037, 3FHL J2104.5+2117 for the W$^+$W$^-$ and $\tau^+\tau^-$ annihilation channels, respectively.}

\begin{figure*}[!ht]
\vspace{-0.5cm}
\includegraphics[width=0.45\textwidth]{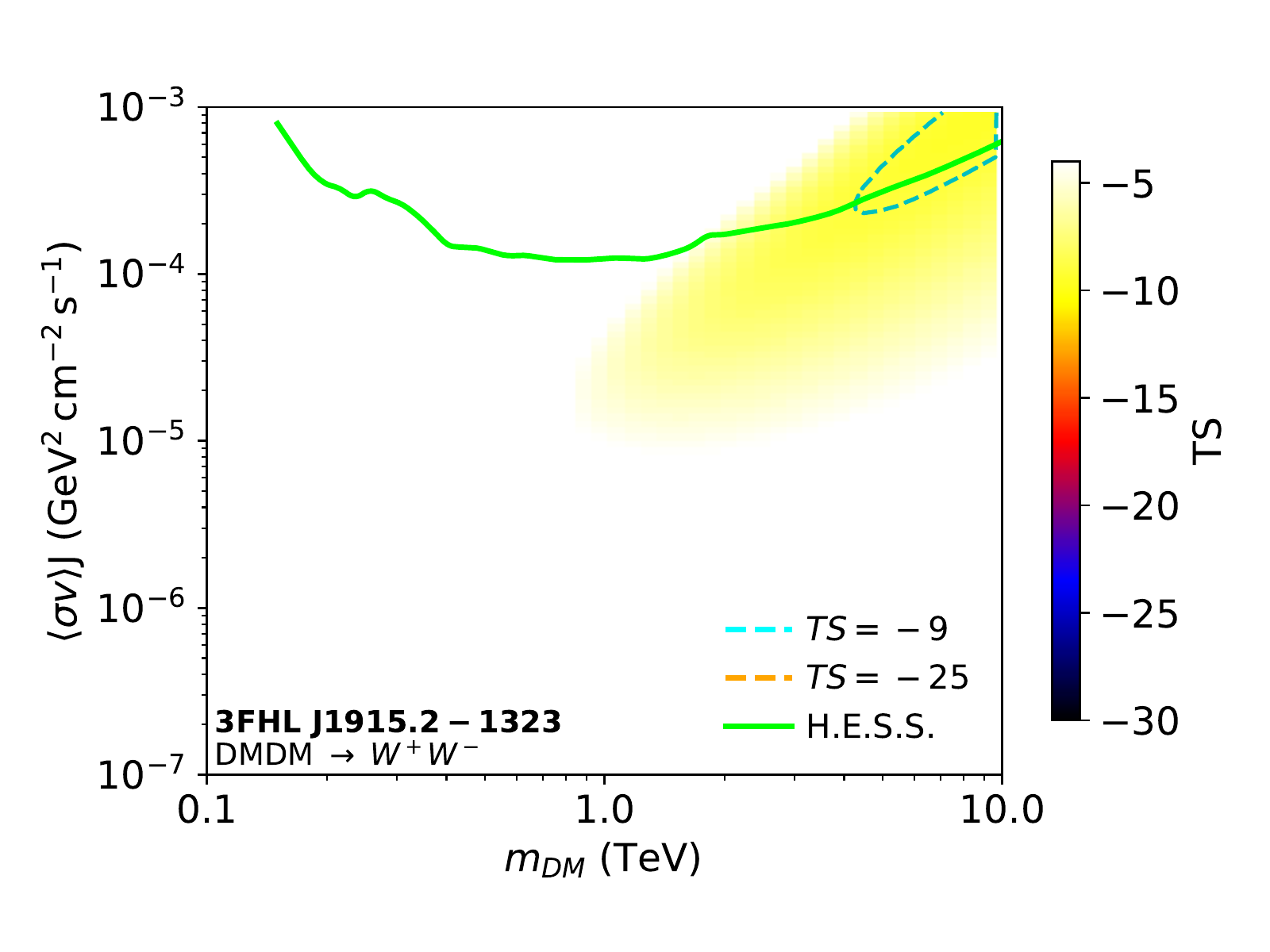}
\vspace{-0.5cm}
\includegraphics[width=0.45\textwidth]{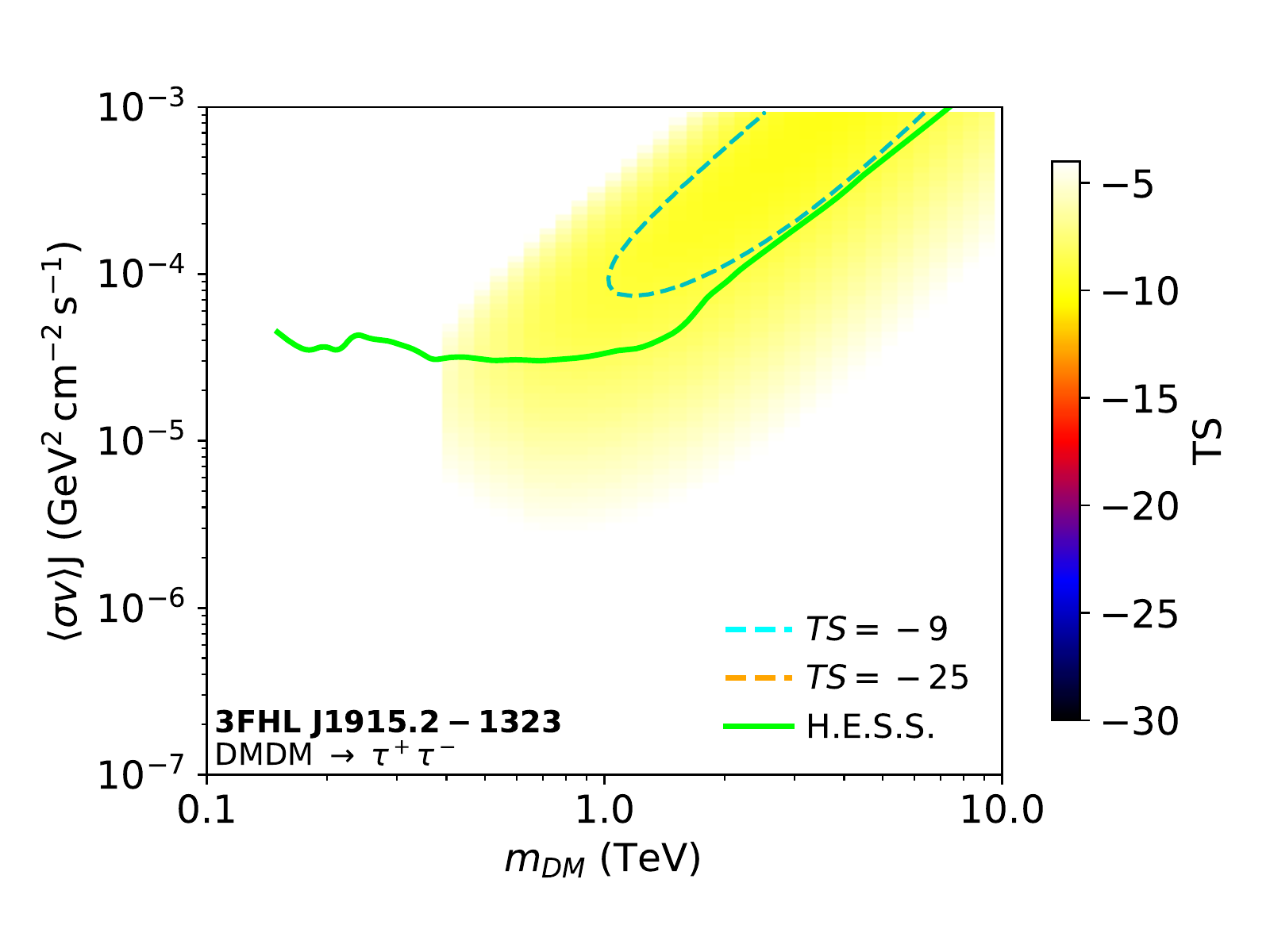}
\includegraphics[width=0.45\textwidth]{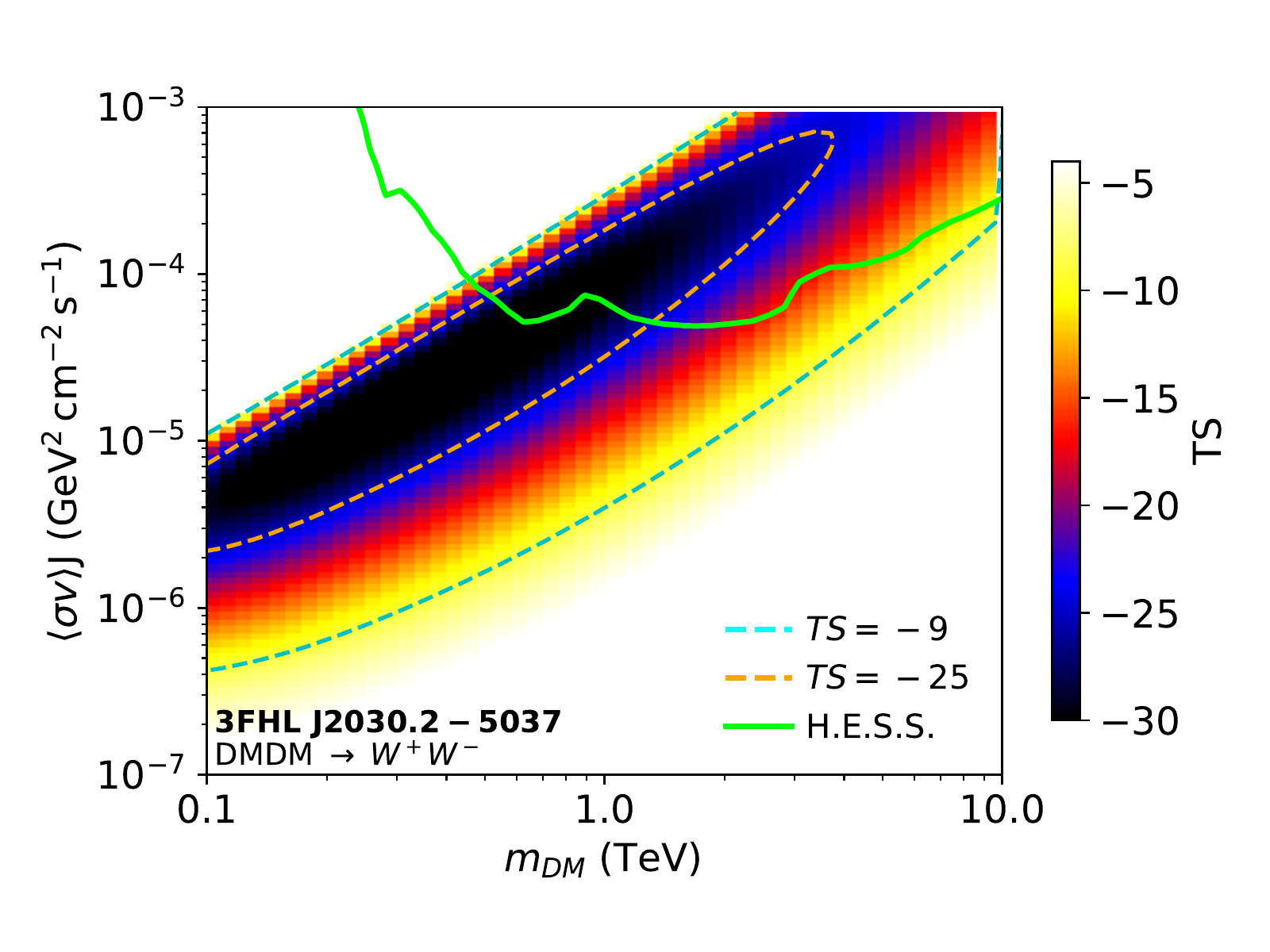}
\includegraphics[width=0.45\textwidth]{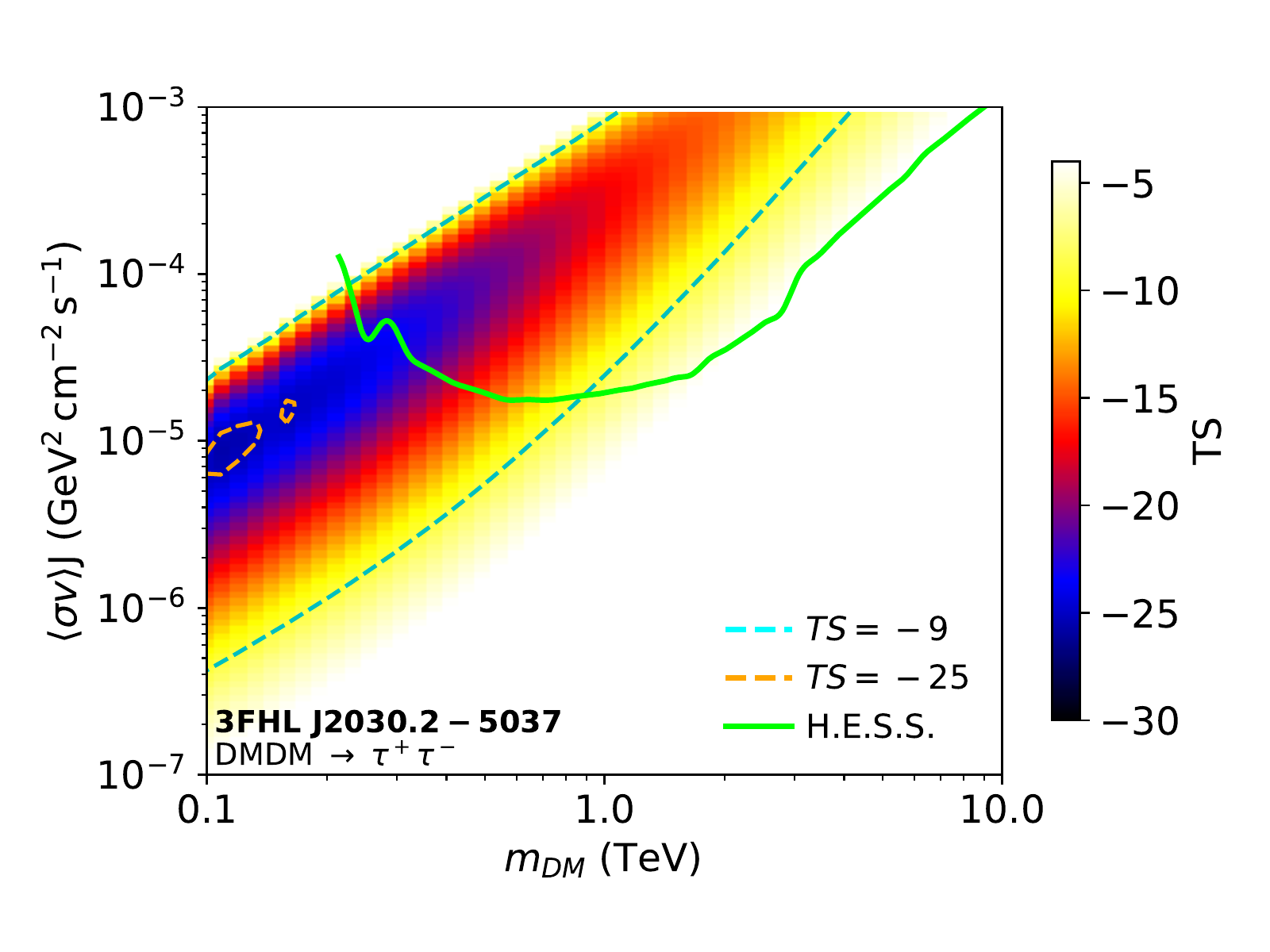}
\includegraphics[width=0.45\textwidth]{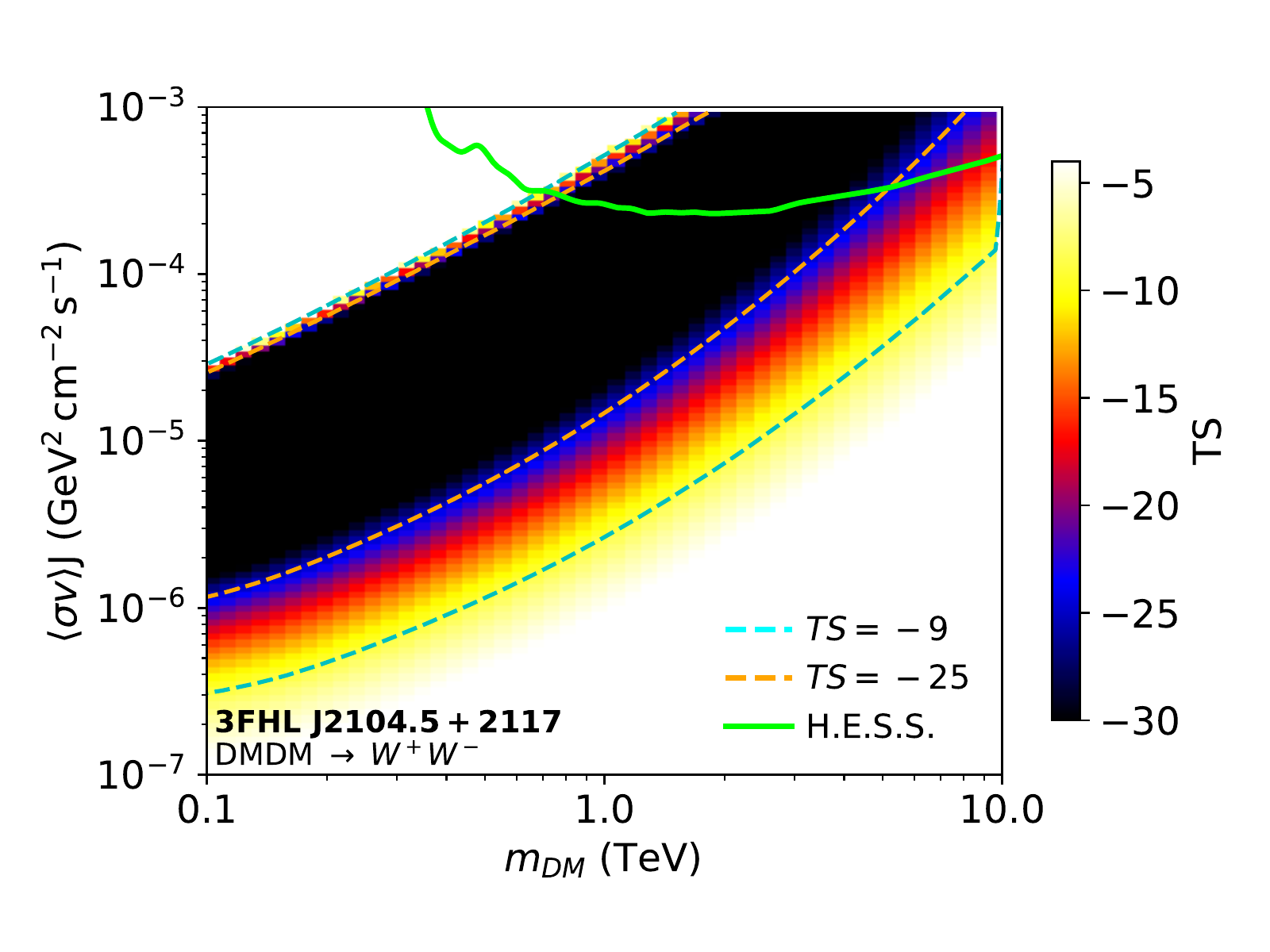}
\hspace{+1.6cm}
\includegraphics[width=0.45\textwidth]{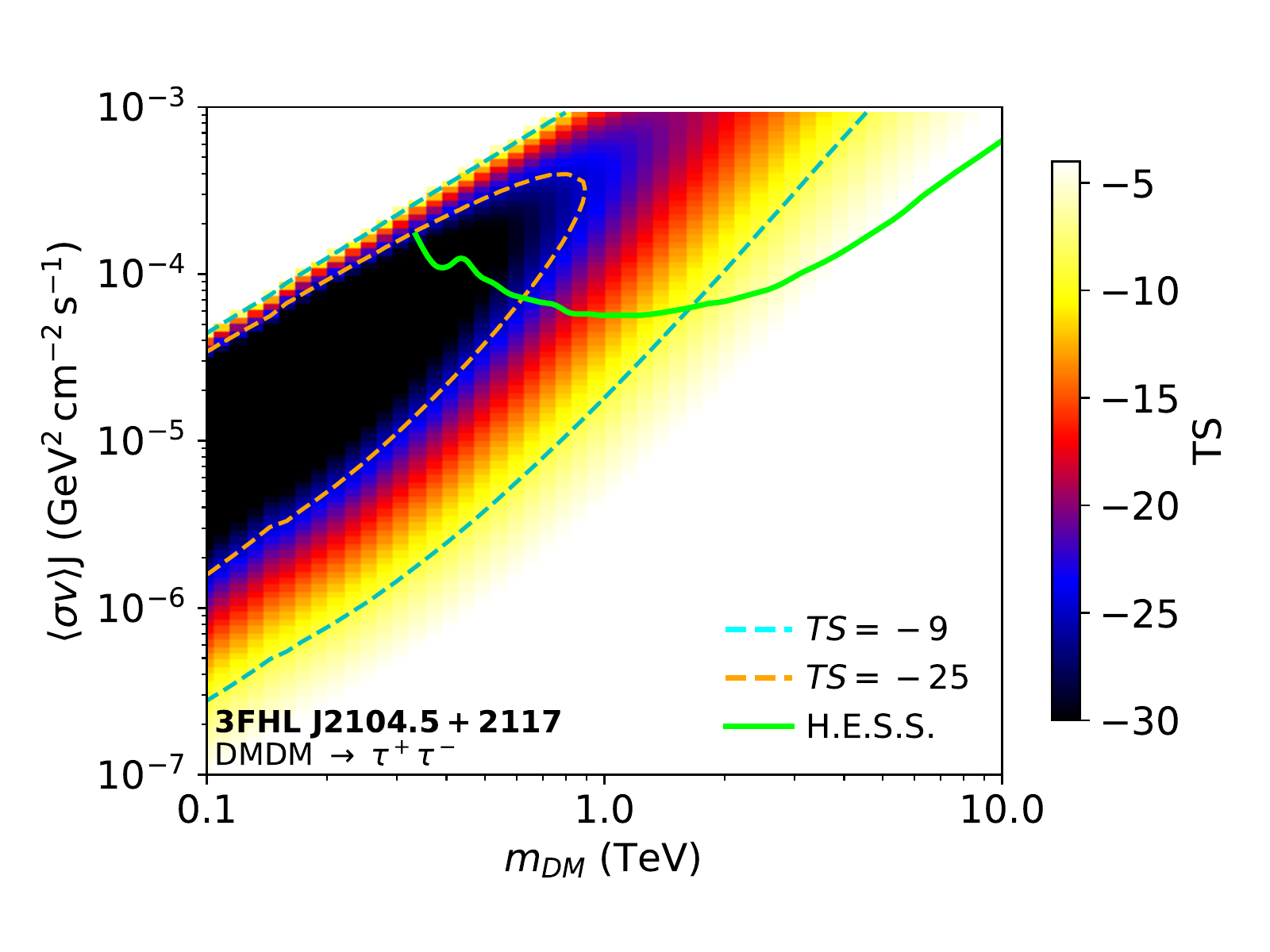}
\caption{Contours of $TS$ computed from \flat datasets on 3FHL J1915.2-1323 (upper panels), 
3FHL J2030.2-5037 (middle panels), 3FHL J2104.5+2117 (lower panels),
 respectively. The contours are given in the ($\langle \sigma v \rangle J$,m$_{\rm DM}$) plane for the $W^+W^-$ (left panels) and $\tau^+\tau^-$ (right panels) annihilation channel. The cyan and orange dashed lines show the $-$9 and $-$25 $TS$ contours.
 Overlaid (solid green line) are the H.E.S.S. constraints displayed at 95\% C.L.}
\label{fig:FermiTSmap1}
\end{figure*}

\end{document}